\documentclass{article}
\usepackage{amsfonts,amssymb,amsthm,amsmath}
\usepackage{hyperref}
\usepackage[dvips]{graphics}
\usepackage{graphicx}
\usepackage{color}

\newcommand{\rom}[1]{\mathrm{#1}}

\renewcommand{\simeq}{\cong}

\def\cA{\mathcal{A}}

\def\cF{\mathcal{F}}

\def\cH{\mathcal{H}}

\def\cL{\mathcal{L}}
\def\cM{\mathcal{M}}
\def\cN{\mathcal{N}}

\def\cP{\mathcal{P}}
\def\cQ{\mathcal{Q}}

\def\cV{\mathcal{V}}

\definecolor{AV}{rgb}{0.65,0.0,0}
\definecolor{Ella}{rgb}{0.3,0,0.65}
\definecolor{Green}{rgb}{0,0.5,0}

\def\eps{\epsilon}

\def\half{\frac{1}{2}}

\def\no{\nonumber}
\def\beq{\begin{eqnarray}}
\def\eeq{\end{eqnarray}}

\def \RR{{\mathbb{R}}}

\def\mf{\mathfrak}

\def\o{\omega}

\def\be{\begin{equation}}
\def\ee{\end{equation}}
\def\bea{\begin{eqnarray}}
\def\eea{\end{eqnarray}}

\def\Aone{\cA_{(1)}^1}
\def\Atwo{\cA_{(1)}^2}
\def\Azero{\chi_1}
\def\Fone{\cF_{(2)}^1}
\def\Ftwo{\cF_{(2)}^2}
\def\Fzero{\cF_{(1)}}
\def\axone{\chi_2}
\def\axtwo{\chi_3}
\def\fone{F_{(1)}^1}
\def\ftwo{F_{(1)}^2}
\def\gauge{A_{(1)}}
\def\strength{F_{(2)}}
\def\nn{\nonumber}

\begin{document}
\pagestyle{myheadings}
\markboth{\textsc{\small }}{%
  \textsc{\small $G_2$ Dualities in $D=5$ Supergravity and Black Strings }} \addtolength{\headsep}{4pt}

\begin{flushright}
 ULB-TH/09-07\\
\texttt{0903.1645 [hep-th]}
\end{flushright}

\begin{centering}

  \vspace{1.0cm}

  \textbf{\Large{$G_2$ Dualities in $D=5$ Supergravity and Black Strings \\ \vspace{0.5cm}  }}

  \vspace{1.0cm}

  {\large Geoffrey~Comp\`{e}re$^\natural$ , Sophie~de Buyl$^\natural$, Ella Jamsin$^\Diamond$, and Amitabh Virmani$^\Diamond$ }

  \vspace{1.5cm}

\begin{minipage}{.9\textwidth}\small \it \begin{center}$^\natural$
University of California at Santa Barbara \\
CA--93106 Santa Barbara, United States \\
$ \, $ \\
$\Diamond$    Physique Th\'eorique et Math\'ematique,  Universit\'e Libre de
    Bruxelles\\and\\ International Solvay Institutes\\ Campus
    Plaine C.P. 231, B-1050 Bruxelles,  Belgium \\
    \vskip 0.5cm
\texttt{gcompere, sdebuyl@physics.ucsb.edu;  ejamsin, avirmani@ulb.ac.be}\\
    \end{center}
\end{minipage}
\end{centering}
\vspace{1.cm}
\begin{center}
  \begin{minipage}{.9\textwidth}
   \begin{center} \textbf{Abstract}\\\vspace{10pt}\end{center}
Five dimensional minimal supergravity dimensionally reduced on two commuting Killing directions gives rise to a $G_2$ coset model.
The symmetry group of the coset model can be used to generate new solutions by applying group transformations on a seed solution.  We show that on a general solution the generators belonging to the Cartan and nilpotent subalgebras of $G_2$ act as scaling and gauge transformations, respectively. The remaining generators of $G_2$  form a $\mf{sl}(2,\RR) \oplus \mf{sl}(2,\RR)$ subalgebra that can be used to generate non-trivial charges. We use these generators to generalize the five dimensional Kerr string in a number of ways. In particular, we construct the spinning electric and spinning magnetic black strings of five dimensional minimal supergravity. We analyze physical properties of these black strings and study their thermodynamics. We also explore their relation to black rings.
  \end{minipage}
\end{center}
\vfill
\noindent \mbox{}
\raisebox{-3\baselineskip}{%
  \parbox{\textwidth}{ \mbox{}\hrulefill\\[-4pt]}} {
  }
\thispagestyle{empty} \newpage
\tableofcontents
\section{Introduction}

The story of hidden symmetries in gravitational theories dates back to the discovery of Ehlers that, upon dimensional reduction on a circle, four dimensional general relativity possesses an $SL(2,\RR)$ invariance \cite{Ehlers}.\footnote{When combined with the Matzner--Misner group, it leads to an infinite--dimensional symmetry --- the Geroch group --- acting on solutions of Einstein's equations with two commuting Killing vectors (axisymmetric stationary solutions) \cite{Geroch:1970nt}. The Geroch group has been identified with the affine extension of $SL(2,\RR)$, namely the affine Kac--Moody group $SL(2,\RR)^+$. In this paper, we restrict ourselves to \emph{finite} dimensional hidden symmetries.} Since then, the notion of hidden symmetries has been generalized to many gravitational theories in various dimensions. The remarkable discovery of  $E_{7(7)}/SU(8)$ coset describing the scalar sector of $N=8$, $D=4$ supergravity \cite{Cremmer:1978km, Cremmer:1979up} led to an earnest exploration of hidden symmetries for supergravity theories \cite{Julia}. It soon became clear that a large number of supergravity theories reduce to  gravity and p-forms coupled to non-linear sigma models upon dimensional reduction. Such sigma models are maps from a lower dimensional base space to a target space. The target space is generally a coset $G/H$, where $G$ is the group of global isometries of the target space, and $H$ is a subgroup of $G$. The symmetry group of a coset model can be used to generate
new solutions by applying a group transformation to a coset representative of a seed solution. During the second superstring revolution these solution generating techniques were  used extensively to generate a rich spectrum of black holes in string theory (see e.g., \cite{Horowitz:1992jp,Youm:1997hw,Peet:2000hn} for reviews).

In this paper we explore these solution generating techniques in the context of minimal supergravity in five dimensions.
Minimal supergravity in five dimensions is the simplest supersymmetric extension of vacuum gravity. The bosonic sector of the theory is the  Einstein-Maxwell theory with a Chern-Simons term. This theory also arises as a consistent truncation of eleven dimensional supergravity. As a result, supersymmetric and near supersymmetric black holes of five dimensional supergravity admit microscopic interpretation in terms of intersecting M-branes.

The discovery of black rings \cite{Emparan:2001wn, Emparan:2004wy, Elvang:2004rt} (see \cite{Emparan:2006mm, Emparan:2008eg} for reviews and further references)  has attracted renewed interest in exact solutions of this theory.  A five parameter family of black ring solutions characterized by mass, two angular momenta, electric charge, and dipole charge
is conjectured to exist in  minimal supergravity \cite{Elvang:2004xi}.  At present, though, all known smooth black rings have no more than three independent parameters \cite{Elvang:2004rt, Elvang:2004xi, Pomeransky:2006bd}.  The three parameter family in \cite{Elvang:2004xi}  was constructed using boosts and  string dualities, whereas the three parameter family of \cite{Pomeransky:2006bd} was constructed using inverse scattering methods. The solutions of \cite{Elvang:2004xi, Pomeransky:2006bd} do not admit any non-trivial supersymmetric limit to the BPS black ring \cite{Elvang:2004rt}. It is likely that efficient solution generating techniques, like
the one explored
in this paper,  would allow one to construct the most general black ring that will describe thermal excitations above the supersymmetric ring.

The solution generating technique we investigate in this paper is based on the hidden symmetry arising upon
dimensional reduction of five dimensional supergravity down to three dimensions. The resulting theory is
three dimensional gravity coupled to a non linear sigma model.
The sigma model is
 globally invariant under the lowest rank exceptional Lie group $G_{2 (2)}$ \cite{Cecotti:1988qn, Mizoguchi:1998wv, Cremmer:1999du, Cremmer:1997ct, Cremmer:1998px} \footnote{ $G_{2(2)}$ is the maximally split real form of $G_2$. Since it is the only real form of $G_2$ that is relevant for our purposes,  we denote $G_{2(2)}$ simply by $G_2$. At the level of Lie algebras, we denote the maximally split real form of $\mf{g}_2$, often written as $\mf{g}_{2(2)}$, simply by $\mf{g}_2$.}.
 The target space of the sigma model depends on the signature of the
three dimensional base space:
 it is $G_{2}/SO(4)$ for the Lorentzian signature or $G_{2}/(SL(2,\RR)\times SL(2,\RR))$ for the Euclidean signature.

A detailed study of the coset model solution generating techniques for this theory was also performed in \cite{Bouchareb:2007ax} (see \cite{Clement:2008qx,Gal'tsov:2009kb} for reviews). The formalism of \cite{Bouchareb:2007ax} was used in \cite{Tomizawa:2008qr} to generate a new rotating charged Kaluza-Klein black hole solution, {and in \cite{Tomizawa:2009ua} to establish a uniqueness theorem for charged rotating black holes in this theory}. The gravitational subsector of minimal five dimensional supergravity was analyzed in \cite{Giusto:2007fx,Ford:2007th} and the extended $U(1)^3$ five dimensional supergravitiy was treated in \cite{Gal'tsov:2008nz,Gal'tsov:2008sh}. Supersymmetric solutions of both gauged and ungauged five dimensional supergravities were studied using the $G_2/SO(4)$ sigma model in \cite{Berkooz:2008rj}.
Our approach is complementary to theirs in a number of ways. In \cite{Bouchareb:2007ax} a derivation of the three dimensional sigma model was given
 that had the advantage of being more transparent for generating solutions, though the $G_2$ symmetries were not immediately evident. The $G_2$ symmetries were made manifest by solving appropriate Killing equations on the coset manifold
(see also  \cite{Clement:2007qy}). The resulting symmetry transformations were then interpreted  through their action on the three dimensional fields. In this paper we use the derivation of the coset model performed in \cite{Cremmer:1999du, Cremmer:1997ct, Cremmer:1998px}, where the  $G_2$ symmetries are manifest from the beginning.  The originality of our approach  lies in decomposing at the outset the symmetry generators  of $G_{2}$ in three different subalgebras: nilpotent, Cartan, and pseudo-compact generators\footnote{When the dimensional reduction is performed along spacelike Killing vectors only,  this decomposition is identical to the Iwasawa decomposition. The definition of pseudo-compact generators is given in section~\ref{sec:LagrDECOMP}.}. We show that only the pseudo-compact generators generate non-trivial charges. Furthermore, using the prescription of \cite{Giusto:2007fx}, we show that all pseudo-compact generators also preserve the Kaluza-Klein asymptotics. This motivates us to focus our study mainly on the charging transformations in the context of black strings, while we only mention some results for asymptotically flat black holes and black rings.

Our main results can be summarized as follows:
\begin{itemize}
\item We show that the action of the Cartan and nilpotent subalgebras on a general seed solution amounts to scaling and gauge transformations, respectively.
\item  We study in detail how the pseudo-compact generators generalize the metric of the Kerr string.
 We identify each generator with a charging transformation.
 \item Using these transformations we construct: $(i)$ a spinning electrically charged black string  where the electric charge is uniformly smeared over the string direction, and $(ii)$ a spinning magnetic one brane. These solutions were also constructed recently in string theory in \cite{Tanabe:2008vz} using boosts and string dualities. The $G_2$ generating  technique is more efficient for finding these solutions in minimal supergravity.
\item We present an analysis of physical properties and thermodynamics of these black strings.
\item We  explore in some detail the relation between these black strings and black rings. These black strings describe, respectively, the infinite radius limit of the yet to be found  doubly spinning   electrically charged  black ring and doubly spinning dipole black ring.
\end{itemize}
Thanks to the efficiency of the $G_2$ method, the black string describing the infinite radius limit of the most general black ring of five dimensional minimal supergravity can also be constructed. The solution and its thermodynamics will be presented
in a separate publication \cite{monster}.

The rest of the paper is organized as follows. We start with a brief overview of the coset model solution generating technique for four dimensional gravity in section \ref{warm} emphasizing the role of the Iwasawa decomposition.
The dimensional reduction of five dimensional minimal supergravity from five to three dimensions is performed in section \ref{sec:LagrDECOMP} and the resulting non-linear sigma model is presented. For ease of reference some basic facts about $G_2$ are collected at the beginning of section \ref{sec:LagrDECOMP}. A concluding ``recipe'' for generating new solutions using $G_2$ dualities is given at the end of section \ref{sec:LagrDECOMP}. In section \ref{cartannilpotent} we show that the action of the Cartan and nilpotent subalgebras on a general seed solution amounts to scaling and gauge transformations, respectively. In section \ref{solgen} we study the action of the pseudo-compact generators on a variety of solutions of interest: black holes, black rings, and black strings. Our main focus is on black strings.
We construct the spinning electric and spinning magnetic black strings of five dimensional supergravity. In this section we also present an analysis of physical properties of these black strings and study their thermodynamics. Finally, we close with a brief discussion in section \ref{disc}. The details of the 7 $\times$ 7 representation of $G_2$ that we use are relegated to appendix \ref{expl_rep}.
In appendix \ref{Killsym} we present an argument that under $G_2$ dualities the number of commuting Killing symmetries cannot change. In appendix \ref{asym} we identify the generators of the subgroup $\tilde K$ of $G_2$ that preserve the Kaluza-Klein asymptotics and asymptotic flatness. A detailed dictionary with the results of \cite{Giusto:2007fx} is presented in appendix \ref{giusto}.

\section{Warm-up: duality in four dimensional gravity}
\label{warm}
We start by summarizing the main ideas of the solution generating  technique based on coset models  using the simple example of four dimensional vacuum gravity reduced on a timelike Killing direction (see \cite{Pope} for a pedagogical review). The standard reduction from four to three dimensions takes the form
\begin{equation}
ds^2 = -e^{-\phi}(dt+ \cA)^2+e^{\phi}ds^3_{(3)}~,
\end{equation}
where the scalar dilaton $\phi$, the one-form potential ${\cal A}$, and the base metric $ds^2_{(3)}$ only depend on three spatial coordinates. The reduced Einstein equations can be derived from three dimensional gravity coupled to a non-linear sigma model \cite{Ehlers}
\be
\cL = \sqrt{^{(3)}g} \left( ^{(3)}R -
\frac{1}{2}(\partial \phi)^2 - \frac{1}{2}e^{2\phi}(\partial \chi)^2
\right)~.
\ee
The sigma model consists of two scalar fields $(\phi, \chi)$, where the so-called `twist potential' or `axion' $\chi$  is related to the 1-form ${\cal A}$ by the three dimensional Hodge dualization {of the field strength $\cF:=d{\cal A}$},
\be
d\chi = \star (e^{-2\phi} \cF) =
\frac{1}{2}\sqrt{^{(3)}g}\eps_{\alpha\mu\nu}e^{-2\phi}\cF^{\mu\nu}dx^\alpha~.
\ee
 The target space of the coset model is $SL(2,\mathbb{R})/SO(2)$. As a result, the reduced Einstein equations are invariant under the Ehlers group $SL(2,\mathbb{R})$ acting transitively on the target space as an isometry. This $SL(2,\mathbb{R})$ action can be used as a solution generating technique \cite{Ehlers}. Starting with a solution of general relativity one first constructs a set of scalars $(\phi, \chi)$. Then, by acting with an element of the isometry group one finds a transformed set of scalars $(\phi', \chi')$. Dualizing back the new twist potential $\chi'$ to the one form ${\cal A}'$ one obtains a new solution of general relativity.

A very convenient way to systematically classify the action of the $SL(2,\RR)$ transformations is to consider the Iwasawa decomposition of the corresponding Lie algebra $\mf{sl}(2,\RR)$; that is to choose as generators the following combinations of Chevalley-Serre generators $\{h,e,f\}$:
\beq
\qquad h\, , \qquad e \,, \qquad e-f,
\eeq
where $h$ generates the Cartan subalgebra $\mf{h}$, $e$ the nilpotent subalgebra $\mf{n}_+$, and $e-f$ the maximal compact subalgebra $\mf{k}=\mf{so(2)}$.

These three subalgebras act on solutions in the following way:
\begin{itemize}
\item $\mf{h}$: scaling transformation: $\phi \rightarrow \phi + \mu_s,\, \chi \rightarrow e^{-\mu_s}\chi $
\item $\mf{n}_+$: gauge transformation: $\chi \rightarrow \chi + \mu_g $, and
\item $\mf{k}$: proper Ehlers transformation: $(\chi - i e^{-\phi})^{-1} \rightarrow (\chi - i e^{-\phi})^{-1} + \mu_{e},$ which generates the Taub--NUT charge from the Schwarzschild metric.
\end{itemize}
Below we will see that a decomposition resembling the Iwasawa decomposition of $G_2$ plays a very similar role in five dimensional minimal supergravity. A general element of the Cartan subalgebra $\mf{h}$ of $\mf{g_2}$ acts as a scaling transformation, while elements of the nilpotent subalgebra $\mf{n}_+$ act as gauge transformations.  The most interesting generators belong to what we call the pseudo-compact algebra. When the dimensional reduction is performed over one timelike and one spacelike direction, the pseudo-compact algebra is  $\mf{sl}(2,\RR) \oplus \mf{sl}(2,\RR)$. As we will see below, the generators of the pseudo-compact algebra generate non-trivial charges, including for example the five dimensional electric charge.

\section{Dualities in five dimensional minimal supergravity}
\label{sec:LagrDECOMP}

\subsection{Generalities on $G_2$}
\label{genG2}

In order to  fully appreciate the $G_2$ duality of five-dimensional minimal supergravity, some basic facts about this Lie group are needed. For the purpose of reference, we collect here the results we need. For further details we refer the reader to standard references, such as \cite{Humphreys:1980dw}.

The algebra $\mf{g_2}$ is the smallest of the exceptional Lie algebras. Its rank is $2$ and and its dimension is $14$. Its Dynkin diagram is presented in Figure 1.
\begin{figure}[h!]
\begin{center}
\includegraphics[width=4cm]{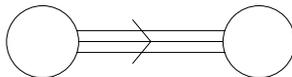}\label{fig:dynkin}
\caption{Dynkin diagram of $G_2$.}
\end{center}
\end{figure}

{\color{black}
\noindent Each node of this diagram corresponds to a triple of Chevalley generators $\{H_a, E_a, F_a\}, a=1,2$.
The $H_a$'s   span  the Cartan subalgebra $\mf{h}$ of $\mf{g_2}$.
The $E_a$'s are the generators associated to the two simple roots $\vec\alpha_1$ and $\vec\alpha_2$ of $\mf{g_2}$. These generators satisfy the Chevalley relations
\bea
\left[ H_1, E_1\right] &=& 2 E_1~, \qquad \left[H_2, E_1\right] = - 3 E_1~, \qquad  \left[E_1, F_1\right] = H_1~, \nn \\
\left[H_1, E_2\right] &=& -E_2, \qquad \left[H_2, E_2\right] = 2 E_2~, \qquad \ \  \ \left[E_2, F_2\right] = H_2~.
\eea
The simple roots belong to the dual $\mf{h^\star}$ of $\mf{h}$.

By taking multiple commutators of $E_a$'s, and using Serre relations, one obtains a set of four more positive generators $E_k,\, k = 3, \ldots, 6$.
More explicitly, one can take them to be
\beq
E_3 = [E_1,E_2]\, ,  \hspace{1cm} E_4 = [E_3,E_2]\, ,  \hspace{1cm} E_5 = [E_4,E_2]\, ,  \hspace{1cm} E_6 =
[E_1, E_5]\,  . \label{genepos}
\eeq
%
%
%
The set of the six positive generators $E_j, j=1,\dots ,6$ form a nilpotent subalgebra $\mf n_+$ of $\mf{g_2}$. To each of these generators corresponds a negative generator $F_j$, associated to the corresponding negative root, and they form another nilpotent subalgebra $\mf n_-$ of $\mf{g_2}$. In the basis
\bea
\label{newbasis}
h_1 &=& \frac{1}{\sqrt{3}} H_2, \quad  h_2 = H_2 + 2 H_1, \nn \\
e_1 &=& E_1, \quad e_2 = \frac{1}{\sqrt{3}}  E_2, \quad e_3 = \frac{1}{\sqrt{3}} E_3,\nn \\
e_4 &=& \frac{1}{\sqrt{12}} E_4, \quad  e_5 = \frac{1}{6} E_5, \quad e_6 = \frac{1}{6} E_6, \nn \\
f_1 &=& F_1, \quad f_2 = \frac{1}{\sqrt{3}}  F_2, \quad f_3 = \frac{1}{\sqrt{3}} F_3, \nn \\
f_4 &=& \frac{1}{\sqrt{12}} F_4, \quad  f_5 = \frac{1}{6} F_5, \quad  f_6 = \frac{1}{6} F_6~,
\eea
the positive roots take the following values:}
\begin{center}
\begin{tabular}{ll}
$\vec\alpha_1 = (-\sqrt{3},1)$,& \quad
$\vec\alpha_2 = (\frac{2}{\sqrt 3},0)$,\\
$\vec\alpha_3 = (-\frac{1}{\sqrt 3},1)=\vec\alpha_1+\vec\alpha_2$,& \quad
$\vec\alpha_4 = (\frac{1}{\sqrt 3},1)=\vec\alpha_1+2\vec\alpha_2$, \\
$\vec\alpha_5 = (\sqrt 3,1)=\vec\alpha_1+3\vec\alpha_2$,& \quad
$\vec\alpha_6 = (0,2)=2\vec\alpha_1+3\vec\alpha_2$.
\end{tabular}
\label{posroots}
\end{center}
The twelve roots of $\mf{g_2}$ are represented in Figure 2.
\begin{figure}[h!]
\begin{center}
\includegraphics[width=4cm]{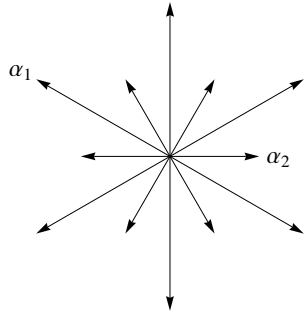}
\label{fig:rootsys}
\caption{The root system of $G_2$.}
\end{center}
\end{figure}

The symmetry between the positive and negative subalgebras $\mf{n_+}$ and $\mf{n_-}$ can be expressed through an involutive automorphism $\tau$, which acts as
\beq
\tau(e_i)=-f_i,\ \  \tau(f_i)=-e_i,\ \ \tau(h_i)=-h_i. \label{chevalley}
\eeq
This automorphism is known as the Chevalley involution of a Lie algebra. The set of elements invariant under the Chevalley involution is the maximal compact subalgebra $\mf{k}$:
\beq
{\mf{k}}=\big\{x\in\mf{g}_2 \,  | \, \tau(x)=x\big\}.
\eeq
It is generated by the elements $\{e_i-f_i\}$, and in the case of $\mf{g_2}$ it is isomorphic to $\mf{so}(4)$. In the study of hidden symmetries, the Chevalley involution and the maximal compact subalgebra are of importance when one compactifies only on spacelike directions.

In the following, we will be interested in compactifying five dimensional minimal supergravity over one spacelike and one timelike Killing direction.
When one first compactifies along a direction of signature $\epsilon_1$ and then along a direction of signature $\epsilon_2$---where $\epsilon_{1,2}$ take values $+1$ or $-1$ depending upon whether the reduction is performed over a spacelike or a timelike direction---, the pertinent involution $\tilde \tau$ is given by the following relations
\begin{eqnarray}
\tilde \tau(h_1) &= &- h_1, \qquad \tilde \tau(h_2)=-h_2, \no \\
\tilde  \tau(e_1) &= & -\eps_1 \eps_2 f_1, \qquad \tilde \tau(e_2) = -\eps_1 f_2,\qquad  \tilde \tau(e_3) =  -\eps_2 f_3, \no \\
\tilde  \tau(e_4) &=&  -\eps_1 \eps_2 f_4,\qquad \tilde \tau(e_5) =  -\eps_2 f_5, \qquad \tilde \tau(e_6) = -\eps_1 f_6 \, . \label{inv}
\end{eqnarray}
Note that for a reduction over two spacelike directions, $\epsilon_1=\epsilon_2=+1$, one finds back the Chevalley involution (\ref{chevalley}). In the case $\epsilon_1=-1$ and $\epsilon_2=+1$, the $\tilde \tau$ involution differs from the Chevalley involution because $\tilde\tau(e_i)=+f_i$ for some generators.

The subalgebra of elements fixed under $\tilde \tau$  is no longer compact. We call it a `pseudo-compact' subalgebra and denote it by $\tilde{\mf{k}}$. It consists of all the elements of the form $\{e_i+\tilde\tau(e_i)\}$, that is,
\beq
k_1 &=& e_1 + f_1\, , \qquad k_2 \, = \, e_2+f_2 \, , \qquad k_3  = e_3-f_3 \, , \no \\
k_4 &=& e_4 + f_4 \, , \qquad k_5 \, =\,  e_5 - f_5\, , \qquad k_6 = e_6 + f_6  \, .\label{eq:k}
\eeq
These generators generate the $\mf{sl}(2,\RR)\oplus\mf{sl}(2,\RR)$ algebra.
An easy way to see this is to rewrite them in the new basis as follows
\beq
k_h &=& \half (k_1 + \sqrt{3} k_4)\, , \qquad \bar k_h = \frac{1}{2}(3 k_1 - \sqrt{3} k_4) \\
k_e &=& { \sqrt{3} \over 4}(k_2 + k_3) +{1 \over 4} (k_5 + k_6) \, , \qquad \bar k_e =\frac{\sqrt 3}{4}(k_2+k_3)-\frac{3}{4}(k_5+k_6) \no \\
k_f  &=& {\sqrt{3}\over 4}(k_2 - k_3) +{1 \over 4} (k_6 - k_5) \, , \qquad \bar k_f = \frac{\sqrt 3}{4}(k_2-k_3)+\frac{3}{4}(k_5-k_6)  \no \, .
\eeq
In this basis we recognize the usual $\mf{sl}(2,\RR)$ commutation relations for the unbarred generators
\beq
 [k_h,k_e] &=& 2 k_e \, , \qquad [k_h,k_f ] = -2 k_f \, , \qquad [k_e,k_f] = k_h~.
\eeq
It can be easily checked that the barred generators also satisfy  the $\mf{sl}(2,\RR)$ commutation relations,
while the commutators between the unbarred and barred generators vanish.

\subsection{Dimensional reduction from five to three dimensions}
\label{dimred}
The dimensional reductions of five dimensional minimal supergravity to three dimensions were first studied by Cremmer, Julia, Lu, and Pope \cite{Cremmer:1998px}, and by Mizoguchi and Ohta \cite{Mizoguchi:1998wv}. When the reduction is performed over two spacelike Killing directions one obtains three dimensional Lorentzian gravity coupled to the $G_{2}/SO(4)$ coset model. On the other hand, when the reduction is performed over one timelike and one spacelike Killing direction one obtains three dimensional Euclidean gravity coupled to the $G_{2}/(SL(2,\RR)\times SL(2,\RR))$ coset model.  In this section, we briefly review this dimensional reduction, treating both cases simultaneously. See~\cite{Hull:1998br, Cremmer:1998em} for details on compactifications along timelike directions.

Five dimensional minimal supergravity contains a metric $g_5$ and a gauge potential $A_{(1)}^5$ whose field strength is $F^5_{(2)}= dA^5_{(1)}$. It is the simplest supersymmetric extension of vacuum five dimensional gravity. The bosonic sector of the theory is the Einstein-Maxwell theory with a Chern-Simons term.  Our starting point is the bosonic part of the Lagrangian, which is given by
\begin{equation}
\cL_5 = R_5 \star 1 - \half\star  F^5_{(2)}\wedge F^5_{(2)}+\frac{1}{3\sqrt 3} F^5_{(2)}\wedge F^5_{(2)}\wedge A^5_{(1)} \, . \label{5dsugra}
\end{equation}
The reduction of the five dimensional metric $g_5$ leads to the following three-dimensional fields: a three-dimensional metric $g_3$, two dilatons $\phi_1$ and $\phi_2$, a scalar $\chi_1$,
and two Kaluza-Klein one-form potentials $\Aone$ and $\Atwo$.  More explicitly, these fields arise from the following dimensional reduction ansatz for the five dimensinal metric
\begin{eqnarray}
ds^2_5 &=& e^{\frac{1}{\sqrt 3}\phi_1+\phi_2}ds^2_3 +\eps_2 e^{\frac{1}{\sqrt 3}\phi_1-\phi_2}(dz_4 + \Atwo)^2\nonumber\\
& &+\eps_1 e^{-\frac{2}{\sqrt 3}\phi_1}(dz_5+\Azero dz_4+\Aone)^2\, ,
\label{metric5}
\end{eqnarray}
where the fields $\phi_1, \, \phi_2, \,  \Aone, \,  \Atwo, \Azero$, and the three-dimensional metric $ds^2_{3}$ do not depend on the $z_4$ and $z_5$ coordinates. One can also think of this reduction as a two step process. The first step being the reduction from five to four dimensions over $z_5$, and the second being the reduction from four to three dimensions over $z_4$. In each step, the reduction can be performed over either a spacelike or a timelike Killing direction.  The sign $\epsilon_{i}$ is $+1$ when the reduction is performed over a spacelike direction, and $-1$ for a timelike direction.
We denote the field strengths associated to  $\Azero$, $\Aone$, and $\Atwo$ by $\Fzero$, $\Fone$, and $\Ftwo$ respectively. They are defined to be,
\begin{eqnarray}
\Fzero&=& d\Azero\, , \nonumber\\
\Fone&=& d\Aone+ \Atwo\wedge d\Azero\, ,\nonumber \\
\Ftwo&=&d\Atwo \, .\nonumber
\end{eqnarray}

The reduction of the five-dimensional gauge potential $A^5_{(1)}$ leads to the three-dimensional gauge potential $\gauge$ and two scalars $\axone$ and $\axtwo$,
\begin{eqnarray}
\label{pot5}
A^5_{(1)} = \gauge+ \axtwo dz_4+ \axone dz_5,
\end{eqnarray}
with associated field strength $\strength$, $\fone$ and $\ftwo$ defined to be,
\begin{eqnarray}
\fone&=& d\axone,\nn\\
\ftwo &=&d\axtwo-\Azero d\axone ,\\
\strength&=& d\gauge- d\axone\wedge (\Aone- \Azero \Atwo)- d\axtwo\wedge \Atwo.\nn
\end{eqnarray}

The reduced Lagrangian in terms of these variables is given by \cite{Cremmer:1999du, Cremmer:1998em}
\begin{eqnarray}
\cL &=& R \star 1- \half \star d\vec\phi\wedge d\vec\phi - \half \eps_1\eps_2 e^{\vec\alpha_1 \cdot \vec\phi} \star  \Fzero\wedge \Fzero- \half  \eps_1 e^{\vec\alpha_2 \cdot \vec\phi}\star  \fone\wedge \fone\nonumber\\
&&- \half \eps_2 e^{\vec\alpha_3 \cdot \vec\phi}\star  \ftwo\wedge \ftwo- \half e^{-\vec\alpha_4 \cdot \vec\phi}\star  \strength\wedge \strength- \half  \eps_1 e^{-\vec\alpha_5 \cdot \vec\phi}\star  \Fone \wedge \Fone \nonumber\\
&&-\half \eps_2 e^{-\vec\alpha_6 \cdot \vec\phi}\star  \Ftwo \wedge \Ftwo
 +\frac{2}{\sqrt 3} d\axone\wedge d\axtwo\wedge \gauge,\label{eq:redLagr}
\end{eqnarray}
where $\vec\phi=(\phi_1,\phi_2)$ and $\vec \alpha\cdot\vec \beta$ is the Euclidean inner product. The six doublets $\vec\alpha_1,\dots,\vec\alpha_6$ correspond precisely to the six positive roots of the exceptional Lie algebra $\mf{g_2}$, given in section \ref{genG2}.

It is clear from the Lagrangian (\ref{eq:redLagr})  that the roots $\vec\alpha_1$, $\vec\alpha_2$ and $\vec\alpha_3$ are respectively associated to the three axions $\Azero$, $\axone$ and $\axtwo$. The other roots come with one-form potentials that in three dimensions can be dualized into scalars. The signs of the exponentials appearing in front of the kinetic terms of the Lagrangian (\ref{eq:redLagr}) indicate whether the field associated to the root is a scalar field ($+\vec\alpha \cdot \vec\phi$) or a one-form potential ($-\vec\alpha \cdot \vec\phi$) that we need to dualize. Note that the signatures of the compactified directions do not appear in the definitions of the field strengths. They only appear as the signs of the kinetic terms in the Lagrangian (\ref{eq:redLagr}).

We now define the axions  $\chi_4$, $\chi_5$, and $\chi_6$  dual to the one forms $\gauge$, $\Aone$, and $\Atwo$. These axions are associated to the roots $\vec\alpha_4,\vec\alpha_5$, and $\vec\alpha_6$.
Recall that in the process of dualisation
the role of the Bianchi identities is interchanged with the role of the equations of motion. Therefore, the easiest way to do the dualisation is to treat the field strengths as fundamental fields (see e.g., \cite{Pope}).  To this end, we  first rewrite the Chern-Simons term of the dimensionally reduced Lagrangian in terms of the field strengths $F_{(2)}$, $\cF^1_{(2)}$ and $\cF^2_{(2)}$ as
\begin{eqnarray}
\text{Chern-Simons} &=& \frac{1}{\sqrt 3}(\chi_2 d\chi_3 - \chi_3 d\chi_2)\wedge \strength+\frac{1}{3\sqrt 3}\chi_2 (\chi_3 d\chi_2-\chi_2 d\chi_3)\wedge \Fone\nonumber \\
&&+\frac{1}{3\sqrt 3}(\chi_3-\chi_1\chi_2)(\chi_3 d\chi_2-\chi_2 d\chi_3)\wedge\Ftwo.
\end{eqnarray}
As the next step we introduce the axions $\chi_4$, $\chi_5$ and $\chi_6$
 as Lagrange multipliers for the Bianchi identities
of the  field strengths  $F_{(2)}$, $\cF^1_{(2)}$, and $\cF^2_{(2)}$. By construction, the variations with respect to the axions give the Bianchi identities. The variations with respect to the field strengths now give purely algebraic equations of motion, which allow us to introduce the dual one-form field strengths $G_{(1)4}$, $G_{(1)5}$ and $G_{(1)6}$ for the three axions:
\begin{eqnarray}
e^{-\vec\alpha_4 \cdot \vec\phi}\star \strength&\equiv &G_{(1)4} = d\chi_4+\frac{1}{\sqrt 3} (\chi_2 d\chi_3 - \chi_3 d\chi_2), \nonumber \\
\eps_1 e^{-\vec\alpha_5 \cdot \vec\phi} \star  \Fone& \equiv &G_{(1)5} = d\chi_5 - \chi_2 d\chi_4 + \frac{1}{3\sqrt{3}} \chi_2 (\chi_3 d\chi_2 - \chi_2 d\chi_3), \label{dualfields}\\
\eps_ 2 e^{-\vec\alpha_6 \cdot \vec\phi}\star  \Ftwo  &\equiv &G_{(1)6} = d\chi_6- \chi_1 d\chi_5+ (\chi_1\chi_2 - \chi_3)d\chi_4\nonumber\\
&& \qquad \quad \: + \:  \frac{1}{3\sqrt{3}} (-\chi_1\chi_2+\chi_3) (\chi_3 d\chi_2 - \chi_2 d\chi_3) \nonumber.
\end{eqnarray}
In terms of the new variables, the Lagrangian becomes
\begin{eqnarray}
\cL &=& R \star 1- \half \star d\vec\phi\wedge d\vec\phi - \half \eps_1\eps_2 e^{\vec\alpha_1 \cdot \vec\phi} \star  d\chi_1 \wedge d \chi_1 - \half  \eps_1 e^{\vec\alpha_2 \cdot \vec\phi}\star  d\chi_2\wedge d\chi_2\nonumber\\
&&- \half \eps_2 e^{\vec\alpha_3 \cdot \vec\phi}\star  (d\chi_3-\chi_1 d\chi_2)\wedge (d\chi_3-\chi_1 d\chi_2) + \half\eps_t e^{\vec\alpha_4 \cdot \vec\phi}\star G_{(1)4} \wedge  G_{(1)4} \nonumber\\
&&+  \half \eps_1 \eps_t e^{\vec\alpha_5 \cdot \vec\phi}\star  G_{(1)5} \wedge  G_{(1)5} +\half\eps_2 \eps_t e^{\vec\alpha_6 \cdot \vec\phi}\star  G_{(1)6} \wedge  G_{(1)6}~,\label{Lcoset_implicit}
\end{eqnarray}
where $\eps_t$ denotes the  signature of the three dimensional metric. It appears in this expression because of the relation $\star \star \omega_{(1)} = \eps_t \omega_{(1)}$ for any one-form $\omega_{(1)}$.

To summarize, the three-dimensional theory is determined by a three-dimensional metric and a set of eight scalar fields: two dilatons $\phi_1$ and $\phi_2$ and six axions $\chi_1,\dots,\chi_6$.

\subsection{The non-linear $\sigma$-model for $G_2 / \tilde{K} $}
\label{sigma}
It turns out that the Lagrangian (\ref{Lcoset_implicit}) can be rewritten as
\beq \cL = R\star 1 + \cL_{coset} \, , \eeq
where $\cL_{coset}$ is the Lagrangian of a non-linear $\sigma$-model for the coset $G_2/\tilde{K}$, with an appropriate subgroup $\tilde{K}$
depending on the signature of the reduced dimensions.  We can write a coset representative $\cV$ for the coset $G_2/\tilde{K}$ in the Borel gauge\footnote{For a discussion of subtleties associated with this gauge choice see \cite{Keurentjes:2005jw, Bossard:2009at}.} by exponentiating the Cartan and positive root generators of $G_{2}$ with the dilatons and axions as coefficients. In order to make contact with our reduced Lagrangian (\ref{Lcoset_implicit}), we do this in the following way
\begin{equation}
\mathbb \cV = e^{\half \phi_1 h_1 + \half \phi_2 h_2} e^{\chi_1 e_1 }e^{-\chi_2 e_2 +\chi_3 e_3}e^{\chi_6 e_6} e^{\chi_4 e_4 -\chi_5 e_5}.
\label{coset}
\end{equation}
This coset representative transforms under a global $G_2$ transformation $g$ and and a local $\tilde{K}$ transformation $k$ as follows:
\beq
\cV\rightarrow k \mathbb \cV g.
\label{transV}
\eeq
A Lie algebra-valued element $v$ can be written using the coset representative in Cartan-Maurer form $d\cV \cV^{-1}$ that decomposes as
\beq
v:= d\cV \cV^{-1}= \cQ + \cP,
\eeq
where $\cQ$ is in $\tilde{\mf{k}}$ and $\cP$ is the projection along the coset.
 $\cQ$ is invariant under the involution $\tilde \tau$ defined in (\ref{inv}) and $\cP$ is anti-invariant under the involution $\tilde \tau$:
 \beq
 \cQ&=&\frac12(v+\tilde\tau(v))~,\\
 \cP&=&\frac12(v-\tilde\tau(v))~.
 \eeq
One can now write a Lagrangian that is manifestly invariant under global $G_2$ and local $\tilde{K}$ as (see e.g., section 9.1 of \cite{Henneaux:2007ej})
\begin{equation}
\cL_{coset} = -\frac{1}{2}\text{Tr}(\star \cP \wedge \cP)\label{Lcoset} \, .
\end{equation}
With the choice of the coset representative (\ref{coset}) the element
$v$ is found to be
\begin{eqnarray}
v=\ d\cV \cV^{-1} &=& \half \phi_1^{\prime }h_1 + \half \phi_2^{\prime }h_2
+e^{\frac{1}{2} \vec\alpha_1 \cdot \vec\phi} \cF_{(1)2}^1e_1
-e^{\frac{1}{2} \vec\alpha_2 \cdot \vec\phi} F_{(1)1}e_2
+e^{\frac{1}{2} \vec\alpha_3 \cdot \vec\phi} F_{(1)2} e_3 \nonumber\\
&&e^{\frac{1}{2} \vec\alpha_4 \cdot \vec\phi} G_{(1)4}e_4
-e^{\frac{1}{2} \vec\alpha_5 \cdot \vec\phi} G_{(1)5}e_5
+e^{\frac{1}{2} \vec\alpha_6 \cdot \vec\phi} G_{(1)6}e_6 ~,
\end{eqnarray}
from which $\cP$ can be readily constructed.
By an explicit calculation one then finds that the coset Lagrangian (\ref{Lcoset}) coincides with the scalar part of the reduced supergravity Lagrangian (\ref{Lcoset_implicit}) for the Cartan involution (\ref{inv}).

\subsection{Acting with $G_2$ in practice}
\label{acting}
Recall that
 our goal is to act on a set of three-dimensional scalar fields $(\phi_1,\phi_2,\chi_1,\dots,\chi_6)$ with an element $G_2$. From the previous section, the way to go seems to be: first construct the coset representative $\cV$ from the scalar fields, and then act as in equation (\ref{transV}) $\cV \rightarrow k \cV g$, where $g$ is a global element of $G_2$ and $k$, the compensator, is a local element of $\tilde{K}$ that must be chosen in order to keep the Borel gauge of $\cV$. In practice,  however,  choosing the right compensator $k$  turns out to be a very difficult task in the cases of most interest.

A much easier way to act on the scalars is provided by the matrix $\cM$,
\beq
\cM := (\cV^{\sharp})  \cV \, , \label{M}
\eeq
where $\sharp$ stands for the \emph{generalized} transposition, which is defined on the generators of $\mf{g}_{2}$ by
\beq
\sharp( x) := - \tilde \tau (x)\, \qquad  \forall \, x \in \mf{g}_{2} .
\label{generalizedtransposition}
\eeq
The matrix $\cM$ transforms under $\cV \rightarrow k \cV g$ in the simple way
\beq
\cM \rightarrow (g^\sharp) \cM g \, . \label{transfM}
\eeq
The new scalars can be extracted from the transformed matrix $\cM$. Note that, the use of the matrix $\cM$ completely avoids the need of constructing the compensator $k$.
In summary, our strategy to find the transformed set of scalars under a $G_2$ action is the following:
\begin{itemize}
\item start with a seed solution of five-dimensional minimal supergravity with two Killing vectors,
\item reduce this solution to three dimensions using the ansatz (\ref{metric5}) and (\ref{pot5}) and dualize the one-forms to obtain a set of eight scalar fields,
\item construct the matrix $\cM$,
\item act on $\cM$ with an element $g$ of $G_2$ as in equation (\ref{transfM}),
\item extract from the new $\cM$ the new scalar fields,
\item uplift back to a five-dimensional solution.
\end{itemize}
In the last step, remember that $\chi_4$, $\chi_5$, and $\chi_6$ are defined in terms of $A_{(1)}$, $\cA_{(1)}^1$, and $\cA_{(1)}^2$ through the duality relations (\ref{dualfields}), and as a consequence, extracting these one-forms requires integrations which may in general be very difficult. Another way to obtain $A_{(1)}$, $\cA_{(1)}^1$, and $\cA_{(1)}^2$, that may be sometimes easier to apply, was described in \cite{Giusto:2007fx} for $SL(3,\RR)/SO(2,1)$ coset model. It can be readily adapted for the case of our interest. The construction proceeds as follows.
We first note that due to the identity
\beq
\cM^{-1}d\cM =2 \cV^{-1} \cP \cV~,
\eeq
the coset Lagrangian (\ref{Lcoset}) can also be written in terms of the matrix $\cM$ as 
{\beq
\cL_{coset} = - {1 \over 8}\mathrm{Tr}(\star(\cM^{-1}d\cM)\wedge(\cM^{-1}d\cM))\, .
\label{cosetLag}
\eeq}
From this form of the Lagrangian it is easy to see that the equation of motion for the matrix $\cM$ takes the form of the conservation of a current
\beq
d\star (\cM^{-1} d\cM ) = 0 \, .
\eeq
As a result, on-shell one can define a new matrix $\cN$ such that
\beq \label{matrixN}
\cM^{-1} d\cM  = \star \, d \cN~.
\eeq
The one-forms $A_{(1)}$, $\cA_{(1)}^1$, and $\cA_{(1)}^2$ can be directly extracted from the matrix $\cN$. Furthermore, the matrix $\cN$ transforms under a global $G_2$ transformation in a simple way
\beq
\cN \longrightarrow  g^{-1}\cN g \label{act2}.
\eeq
Therefore, the transformed one-forms can also be directly  extracted from the transformed matrix $\cN$. In the 7 $\times$ 7  representation of $G_2$ that we use,  given explicitly in appendix \ref{expl_rep},  and for $\eps_1 = - 1$, $\eps_2 = +1$, the relations between the components of the matrix $\cN$ and the one-forms are:
\beq
\cN_{6,1} &=& - \half \cA^2_{(1)} \, , \nn\\
\cN_{5,1} &=& - \half(\cA^1_{(1)} -\chi_1 \cA^2_{(1)} )\, , \nn\\
\label{eqgaugeintermsofN}
\cN_{4,1} &=& {1 \over\sqrt{3} }(A_{(1)} -\chi_2 \cA^1_{(1)} + (\chi_1 \chi_2 - \chi_3)\cA^2_{(1)} )\, .
\eeq
 In the following, we use
both the matrices $\cM$ and $\cN$
to extract the transformed fields.

At this point one would like to understand certain general features (such as the number of commuting Killing symmetries, BPS nature, etc) of the solutions generated using the group action (\ref{transfM}). In appendix \ref{Killsym} we show, for a general finite dimensional coset model, that under a group transformation the transformed solution and the seed solution must have the same number of commuting Killing symmetries.
The question of how the BPS nature of solutions changes under group transformations in three dimensional Euclidean gravity coupled to a coset model is subtle. For many cases of interest the subgroup $\tilde K$ is non-compact, and, as a result, the Iwasawa decomposition does not cover the whole group $G$.  In \cite{Bossard:2009at} it is noted that elements of $G$ that cannot be decomposed into the Iwasawa form map non-BPS solutions to BPS solutions. In this paper we do not deal with any of these subtleties. We exclusively work with elements of $G_2$ that \emph{can} be decomposed in the Iwasawa form (with non-compact $\tilde K$), and we restrict our attention only to non-BPS solutions. A study of how the BPS nature of solutions changes under $G_2$ transformations is left for the future.

\section{Action of the Cartan and nilpotent subalgebras on general solutions}
\label{cartannilpotent}

In the following, we study separately the action of the Cartan, nilpotent, and $\tilde{\mf{k}}$ subalgebras. The actions of the Cartan and nilpotent subalgebras are simple enough that they can be studied on a general set of scalar fields $\phi_1,\phi_2,\chi_1,\dots,\chi_6$, for any type of reduction.
The action of the $\tilde{\mf{k}}$ subalgebra is more involved. It is analyzed on a particular seed solution in the next section.

\subsection{Cartan subalgebra}
Under the action of a general element of the Cartan subalgebra $a_1h_1+a_2h_2$, that is, acting on $\cM$ with $g=e^{a_1h_1+a_2h_2}$ in equation (\ref{transfM}), the scalars transform as
\beq
\phi_i&\rightarrow&\phi_i+C_i(a_1, a_2)~, \nn\\
\chi_j&\rightarrow&K_j(a_1,a_2)\chi_j~,
\eeq
where $C_i(a_1, a_2)$ and $K_j(a_1,a_2)$ are functions of the transformation parameters only. In other words, this transformation shifts the dilatons and dilates the axions. The constants $C_i$ and $K_j$ are such that the five-dimensional metric and gauge field are modified in the following way,
\begin{eqnarray}
ds^2_5 &\rightarrow& e^{\frac{1}{\sqrt 3}\phi_1+\phi_2}\mathrm a^2\,ds^2_3 +\eps_2 e^{\frac{1}{\sqrt 3}\phi_1-\phi_2}(\mathrm b\,dz_4 + \mathrm a\,\Atwo)^2, \nonumber\\
&&+\eps_1 e^{-\frac{2}{\sqrt 3}\phi_1}(\mathrm c\,dz_5+\Azero \mathrm b\,dz_4+\mathrm a\,\Aone)^2\, ,
\end{eqnarray}
\begin{eqnarray}
A^5_{(1)} \rightarrow \mathrm a\,\gauge+ \axtwo\wedge\mathrm  b\,dz_4+ \axone\wedge \mathrm c\,dz_5,
\end{eqnarray}
where $\mathrm a,\mathrm b$ and $\mathrm c$ are also functions of the parameters $a_1$ and $a_2$  alone. Therefore, we conclude that the transformation by a general element of the Cartan subalgebra is equivalent to doing the following scalings
\beq
ds^2_3&\rightarrow&\mathrm a^2\,ds^2_3\nn,\\
\Aone ,\Atwo ,\gauge&\rightarrow&\mathrm a\, \Aone ,\mathrm a\,\Atwo ,\mathrm a\,\gauge,\nn\\
z_4&\rightarrow&\mathrm b\,z_4,\nn\\
z_5&\rightarrow&\mathrm c\,z_5~.
\eeq
In working with rotational Killing fields this transformation generically generates conical singularities, and, as a result, does not always preserve the asymptotic structure of the seed spacetime.

\subsection{Nilpotent subalgebra}
Under the action of a general six parameter element $b_1\, e_1+\dots+b_6 \, e_6\:$ of the nilpotent subalgebra of $\mf{g_2}$, the dilatons are unchanged but the
axions mix among each other. The mixing is such that the combinations of scalars appearing in the expressions for $\Fzero$, $\fone$, $\ftwo$ and in equations (\ref{dualfields}) for $G_{(1)4}$, $G_{(1)5}$ and $G_{(1)6}$ are unchanged. The first three axions transform as
\beq
\chi_1&\rightarrow&\chi_1+b_1~,\nn\\
\chi_2&\rightarrow&\chi_2-b_2~,\nn\\
\chi_3&\rightarrow&\chi_3+b_1\chi_2-\frac{b_1b_2}{2}+b_3~,
\eeq
and it can be easily verified that this leaves $\Fzero$, $\fone$, and $\ftwo$ unchanged. The last three axions transform in a more complicated way, involving non-linear terms, but again it  can be verified that $G_{(1)4}$, $G_{(1)5}$, and $G_{(1)6}$ are left unchanged. As a consequence, the action of the nilpotent subalgebra
is just a (possibly large\footnote{In the present context, a gauge transformation is called large if it acts non trivially on the asymptotic behavior.}) gauge transformation.

\section{Action of $\mf{sl}(2,\RR)\oplus\mf{sl}(2,\RR)$ on black strings}
\label{solgen}

We now turn to the action of the subalgebra $\tilde{\mf{k}}$. We only consider the case $\eps_1 = - 1$, $\eps_2 = +1$. As in the example of four dimensional general relativity of section \ref{warm}, the subalgebra $\tilde{\mf{k}}$ generically generates charges that a seed solution does not carry. This is in contrast with the action of the Cartan and nilpotent subagebras that do not generate non-trivial charges. In appendix \ref{Killsym} we show that under a general $G_2$ transformation the transformed solution and the seed solution must have the same number of commuting Killing symmetries. These transformations can however alter the asymptotic structure of the spacetime. Nevertheless, by studying the way the matrix $\cM$ transforms, we show in appendix \ref{asym} that all the generators of $\tilde{\mf{k}}$ preserve the Kaluza-Klein asymptotics. This suggests that the action of $\tilde{\mf{k}}$ on black strings is very rich. In this section we study in detail its action on a particular seed solution, namely, the five dimensional Kerr string.

The four dimensional Kerr metric is given by
\be
ds^2 = -f (dt + \omega d\phi)^2 + \Sigma \left( \frac{dr^2}{\Delta} + d \theta^2\right) + \frac{\Delta}{f} (1-x^2) d\phi^2~,
\ee
where the metric functions are
\bea
\Sigma = r^2 + a^2  x^2, \qquad f = 1 - \frac{2 m r}{\Sigma}, \qquad \omega = \frac{2 m r}{\Sigma - 2m r} a (1-x^2)  \qquad \Delta = r^2 - 2 m r + a^2.
\label{funcs}
\eea
For calculational convenience we use $x = \cos \theta$ as one of the coordinates, instead of the polar coordinate $\theta$. The variable $x$ lies in the range $-1\le x \le 1$.  For later use we also define the base metric as
\be
ds^2_{base} = \Sigma \left( \frac{dr^2}{\Delta} + d \theta^2\right) + \frac{\Delta}{f} (1-x^2) d\phi^2~.
\label{base}
\ee
By adding a flat direction to the four dimensional Kerr solution we obtain the five dimensional Kerr string. Using the subgroup $\tilde K$ we generalize the metric of the Kerr string in a number of ways.

\subsection{$k_1$: boost}

With the choice of generator $ \sigma \: k_1$, the action on the Kerr string is simply a boost along the string direction
\be
t \rightarrow t \cosh \sigma - z \sinh \sigma~, \qquad z \rightarrow - t \sinh \sigma  + z \cosh \sigma~.
\label{boost}
\ee

\subsection{$k_2$: electric charge}
Acting with the generator  $+ \sqrt{3} \, \alpha_e \, k_2$, the Kerr string gets transformed into a {non-extremal spinning electric string solution.} The electric charge is uniformly smeared over the string direction. The five dimensional metric and gauge field are
\bea
ds^2_{5} &=& - h \: \xi^{-1} f \left( dt + \omega_{\phi} d \phi\right)^2 + h \: ds^2_{base} + \: h^{-2} \, \xi \left[ - \xi^{-1} \beta_t \, f \, (dt + \omega_{\phi} d\phi) + dz\right]^2,  \nonumber \\
A_{\mu} dx^{\mu} &=& \frac{\sqrt{3}}{h} \left( \frac{s}{c} f \, \omega_{\phi} \: d \phi +  s \, c \: (f-1) \: dt +  \frac{c}{s} \, \beta_t \, dz \right),
\label{electric}
\eea
where we have defined the following functions
\bea
h &=& c^2 - s^2 f, \qquad \xi = h^3 - \beta_t^2 f~, \\
\beta_t &=& \frac{2 s^3 m a x }{\Sigma}, \qquad \omega_{\phi} =  \omega c^3~,
\eea
and in order to reduce notational clutter, we have introduced
\be
s := \sinh \alpha_e~, \qquad c :=  \cosh \alpha_e~.
\ee
The rest of the metric functions and the base metric $ds^2_{base}$ are defined in equations (\ref{funcs}) and (\ref{base}) respectively. The static limit of the above solution (i.e., $a\to 0$) is simply
\be
ds^2_{5} = -H^{-2} \tilde f dt^2  + H \left( dz^2 + \frac{dr^2}{\tilde f} + r^2 d \Omega_2^2 \right),
\label{static}
\ee
with
\be
\tilde f = \left( 1 - \frac{2m}{r}\right), \qquad H = 1 + \frac{Q}{r}, \qquad Q = 2 m \, s^2,
\ee
and the gauge field is
\be
A_{t}= - 2 \sqrt{3} \frac{m c s}{r + Q} .
\ee

From the eleven dimensional perspective, we immediately recognize the static solution as representing a configuration of three orthogonal, equally charged, M2 branes wrapped on a six torus (see e.g., \cite{Ar,Peet:2000hn}). The brane configuration is indicated in the following table
\bea
\begin{array}{c c c c c c c c c c c c}
 & t & z_1& z_2& z_3& z_4 & z_5 & z_6 & z & r & \theta & \phi \\
  M2 & \times & \times & \times & - & - & - & - & - &  &  &  \\
  M2 & \times & - & -& \times & \times & - & - & - &  &  &  \\
  M2 &  \times & - & - & - & - & \times & \times  & - &  &  &
\end{array}
\label{M2}
\eea
The spinning solution (\ref{electric}) thus naturally represents the spinning variant of this brane configuration with equal M2 charges. The spinning solution with three \emph{unequal} charges was recently obtained in \cite{Tanabe:2008vz} using  boosts and string dualities. Upon setting the three charges equal, the electric solution of \cite{Tanabe:2008vz} reduces to (\ref{electric})\footnote{The solution given in \cite{Tanabe:2008vz} had typos that we have fixed.}.

We now provide an analysis of physical properties of the boosted generalization of the spinning electric string (\ref{electric}). We study the boosted string configuration because after bending it into a circle and for a specific value of the boost parameter (see equation (\ref{specific}) below), it should give a doubly spinning electrically charged black ring. The boosted string configuration can be obtained by performing a boost (\ref{boost}) with the boost parameter $\sigma$ on (\ref{electric}). We take the new boosted $z$ coordinate to be along an $S^1$ with circumference $2 \pi R$, which allows us to write $z$ in terms of an angular coordinate $\psi$ defined by
\be
\psi = \frac{z}{R}, \qquad 0 \le\psi < 2 \pi. \label{angle}
\ee
Below we use $z$ and $\psi$ interchangeably.

It is easy to see that the solution (\ref{electric}) (as well as the boosted solution) has a regular outer horizon at $r= r_{+} :=  m + \sqrt{m^2 -a^2}$ of topology $R \times S^2$. In addition there is an inner horizon at $r= r_{-} := m - \sqrt{m^2 -a^2}$. The two horizons coincide when $a = m$, which defines the extremal limit.

The ADM stress tensor (see e.g., section 2.1 of \cite{Myers:1999psa}) of the boosted string is
\bea
T_{tt} &=& \frac{m}{2} \left( \cosh^2 \sigma + 1 + 3 s^2 \cosh^2 \sigma  \right)~, \nn \\
T_{zz} &=& \frac{m}{2} \left( \sinh^2 \sigma  - 1 +  3 s^2 \sinh^2 \sigma  \right)~, \nn \\
T_{tz} &=& \frac{m}{2}  (1 + 3 s^2) \sinh \sigma \cosh \sigma~,
\eea
where $T_{tt}$ and $T_{tz}$ are the energy and linear momentum density and $T_{zz}$ is the pressure density of the black string.
Note that the internal spin does not enter in the above stress tensor expressions.
The mass, linear momentum, angular momentum, horizon area, and linear and angular velocities at the outer horizon can be easily calculated. One finds
\bea
M &=&  2 \pi R \, \, T_{tt} = \pi m R \left( \cosh^2 \sigma + 1 + 3 s^2 \cosh^2 \sigma  \right)~, \label{mass_electric} \\
P_{z} &=& 2 \pi R \, \, T_{tz} = \pi m R \, \, (1 + 3 s^2)  \sinh \sigma \cosh \sigma ~, \\
J_{\phi} &=& 2 \pi R \, m a \, c^3 \, \cosh \sigma~,  \\
A_{\rom{H}} &=&  8 \pi^2 R \,  \left(r_{+}^2 + a^2 \right)\, c^3 \cosh \sigma~,\\
v_{z} &=& \tanh \sigma~, \\
\Omega_{\phi} &=& \frac{a}{r_{+}^2 + a^2 } \, \frac{1}{c^3 \cosh \sigma}~. \label{omega_electric}
\eea
The temperature can be calculated from the surface gravity, and the result is
\be
T_{\rom{H}} = \frac{r_{+} - r_{-}}{8 \pi m \, r_{+} \, c^3 \cosh \sigma}.
\ee
As expected, $T_{\rom{H}} = 0$ for the extremal solution with $m = a$.
The total electric charge is
\be
Q_{\rom{E}} =  \frac{1}{16 \pi }\int_{S^{2}\times \RR} \left( \star F - \frac{1}{\sqrt{3}} \: F \wedge A \right)=  \sqrt{3} \, m \, \pi R \: c \,  s \: \cosh \sigma~.
\ee
In addition, for studying thermodynamics, we define ADM tension ${\cal T}$  from the $T_{zz}$ component of the stress tensor
(see e.g., \cite{Townsend:2001rg,
Harmark:2004ch, Kastor:2007wr}), and the potential $\Phi_{\rom{E}}$ from the difference between the values of $A$ at infinity and at the horizon,
\bea
{\cal T} &=& - T_{zz} = \frac{m}{2} \left( 1- \sinh^2 \sigma - 3 s^2 \sinh^2 \sigma  \right), \\
\Phi_{\rom{E}} &=&  - \left( \xi^\mu A_{\mu} |_{\rom{H}} - \xi^\mu A_{\mu} |_{\infty} \right) =   - \sqrt{3} \frac{s}{c \cosh \sigma}~,
\eea
where \be
\xi = \frac{\partial}{\partial t} + \Omega_{\phi} \frac{\partial}{\partial \phi } + v_{z} \frac{\partial}{\partial z }
\ee
is the horizon generating Killing field.

A straightforward calculation using these results then shows that the boosted black string satisfies a Smarr relation
\be
M = \frac{3}{2} \left( \frac{1}{4} A_{\rom{H}} T_{\rom{H}} + \Omega_{\psi} J_{\psi} + \Omega_{\phi} J_{\phi} \right) + Q_{\rom{E}} \Phi_{\rom{E}} +  \frac{1}{2}{\cal T} (2 \pi R)~,
\ee
where we have introduced the `angular' velocity and `angular' momentum \be\Omega_{\psi} = \frac{v_{z}}{R}, \qquad J_{\psi} = P_{z} R. \ee
The first law can also be explicitly verified,
\be
dM = \frac{1}{4} T_{\rom{H}} dA_{\rom{H}} + \Omega_{\psi}d J_{\psi} + \Omega_{\phi}d J_{\phi} +  \Phi_{\rom{E}} d Q_{\rom{E}} + 2 \pi  {\cal T} d R~.
\ee
For the pressureless solution ($T_{zz} =0$) the Smarr relation and the first law are exactly those of \cite{Gauntlett:1998fz}. This hints to the possibility that the pressureless black string correctly describes the infinite radius limit of a five dimensional asymptotically flat black ring. When the electric charge and the internal spin are zero, the Smarr relation and the first law become exactly those of \cite{Kastor:2007wr}.

In the rest of the section, we briefly discuss the action of $k_2$ on other black objects: black holes and black rings. The $k_2$ action generically generates the five dimensional  electric charge, i.e.,  from the eleven dimensional perspective $k_2$ adds three equally charged M2 branes.  It is therefore expected that its action on the doubly spinning Myers-Perry black hole would give rise to the charged rotating non-BPS black hole \cite{Cvetic:1996xz} of five dimensional minimal supergravity. This expectation is indeed realized. A detailed calculation using $G_2$ dualities is already presented in \cite{Bouchareb:2007ax}. We refer the reader to this reference for further details.

The $k_2$ action on black rings is more interesting and more subtle. Recall that the most general non-supersymmetric black rings known so far in minimal supergravity is a three parameter family that has electric charge, dipole charge, two unequal angular momenta, and finite energy above the BPS bound \cite{Elvang:2004xi}\footnote{The Pomeransky Sen'kov solutions \cite{Pomeransky:2006bd} form another three parameter family of non-supersymmetric black rings in minimal supergravity.}. This family has only three independent conserved charges, namely, mass, electric charge, and angular momentum in the ring direction. The dipole charge and the angular momentum on the 2-sphere are not independent parameters. This family was constructed by adding three M2 brane charges using boosts and string dualities on the five dimensional dipole black rings of \cite{Emparan:2004wy}. Using $G_{2}$ dualities we reproduce this calculation by applying $k_2$ on dipole rings of minimal supergravity.

Unfortunately, these solutions do not admit any non-trivial supersymmetric limit to the BPS black ring. A five parameter family of solutions, characterized by mass, two independent angular momenta, electric charge, and dipole charge is conjectured to exist \cite{Elvang:2004xi}. This family would allow to describe thermal excitations above the BPS solution of \cite{Elvang:2004rt}. It is also argued that this family would exhibit continuous non-uniqueness through the dipole charge. In spite of attempts since early on, exact solutions describing such a family remain elusive. It is natural to ask whether one can construct this family using $G_{2}$ dualities.

A significant step forward would be to obtain a smooth doubly spinning electrically charged black ring solution.
One might expect that by applying $k_2$ on Pomeransky Sen'kov solution \cite{Pomeransky:2006bd} one would generate such a configuration. However, this expectation is not realized. As is carefully explained in \cite{Bouchareb:2007ax}, the final solution one gets suffers from Dirac-Misner string singularities. In fact, such Dirac-Misner strings are expected to arise in working with black rings. The reason is as follows:
one can view the $k_2$ action as an efficient way of doing a sequence of boosts and  string-dualities\footnote{The action of $k_2$ on black strings, black holes, and black rings naturally suggests such an interpretation, nevertheless we do not have a precise argument to support this claim. This point deserves further contemplation.}
to generate three M2 brane charges and eventually setting these charges equal. It is well known in the black ring literature \cite{Elvang:2003mj} that one cannot add three independent charges to an otherwise neutral  ring by applying boosts and string dualities. In a certain duality frame, adding the third charge requires applying a boost along the KK direction of a KK-monopole.  Such a boost is incompatible with the  identifications imposed on the geometry by the KK-monopole fibration. Consequently, one ends up generating Dirac-Misner strings.

This difficulty also arises in the construction of \cite{Elvang:2004xi}. There it is sidestepped by starting with a seed solution that has an extra parameter (dipole charge), which can be tuned so that the final solution is free from Dirac-Misner strings. Such an extra parameter is not yet available for the doubly spinning solution.   It was argued in \cite{Bouchareb:2007ax} that, in principle, using $G_2$ dualities it should be possible to generate a six parameter unbalanced black ring, which should lead to a four parameter non-singular electrically charged  black ring.

{
In the absence of the exact ring solution, one can adopt the \emph{blackfold} point of view  \cite{Emparan:2007wm, Caldarelli:2008pz, Camps:2008hb, Emparan:2009cs} (see also \cite{Hovdebo:2006jy}), and consider perturbing the straight boosted black string so as to bend it into a circle of very large radius.
Perturbative construction of black rings is technically challenging in the presence of gauge fields and internal rotations\footnote{To a large extent the considerations of \cite{Emparan:2007wm, Caldarelli:2008pz, Camps:2008hb, Emparan:2009cs} are restricted to neutral singly spinning black rings possibly in an external gravitational potential. The blackfold
methodology is currently being developed for charged branes \cite{Emparan}.}. Furthermore, in our case, it is not guaranteed at the outset, that Dirac-Misner strings will not be generated in bending the pressureless boosted electrically charged black string.
  We will not dwell here on any of the details of such a construction.
  We simply study a boosted version of the electric string (\ref{electric}) as a toy model for a thin doubly spinning electrically charged black ring. The motivation behind such a study comes from the fact that all known smooth black rings with charges \cite{Elvang:2004rt, Elvang:2003yy} and with dipoles \cite{Emparan:2004wy} also become pressureless strings in the infinite radius limit.

The connection between boosted black strings and black rings was first made explicit in \cite{Elvang:2003mj}. The infinite radius limit corresponds to taking the ring radius much larger than the ring thickness, and focusing on the region near the ring. The absence of pressure in this limit reflects the delicate balance of gravitational tension\footnote{The gravitational attractive force appears only at a subleading order in the inverse radius \cite{Elvang:2006dd}.}, electromagnetic interactions, and centrifugal repulsion that the balanced ring represents.
In our case the pressureless condition $(T_{zz} = 0)$ translates into a specific value for the boost parameter
\be
\sinh^2 \sigma = \frac{1}{1 + 3 s^2}. \label{specific}
\ee
Note that when $s\neq 0$ the boost is smaller than in the neutral case. This observation is easily interpreted by noting that sections of the ring at diametrically opposite ends, $\psi$ and $\psi + \pi$, have electric charges of the same sign and therefore they repel each other via the 2-form field strength $F_{\mu \nu}$. As a result, a smaller centrifugal repulsion is needed in order to achieve the mechanical equilibrium. Substituting (\ref{specific}) in equations (\ref{mass_electric})--(\ref{omega_electric}), one can extract certain important information about the balanced doubly spinning electrically charged black ring.

At this point it is useful to understand the relationship between the infinite radius limit of the singular black ring of \cite{Bouchareb:2007ax} and our electric string. Simply taking the infinite radius limit of the solution of \cite{Bouchareb:2007ax} does not yield our pressureless electric string.  The limit corresponds to
\be
k \to \infty~, \qquad  \nu \to  \frac{a^2}{2k^2}~, \qquad y \to -\frac{\sqrt{2} k}{r}~, \quad \mbox{and} \quad \lambda \to \frac{\sqrt{2} m}{k}.
\ee
The factors of $\sqrt{2}$ are necessary in order to get the standard normalization of the final coordinates in the electric charge going to zero limit. We find that the electric string obtained from \cite{Bouchareb:2007ax} black ring in this limit is pathological.}

\subsection{$k_3$: generates nothing}
$k_3$ does not do anything on the Schwarzschild string, Kerr string, and NUT string. To get the final metric in exactly the original coordinates, certain gauge transformations are required.

\subsection{$k_4$: magnetic charge}
Applying a $k_4$ transformation with the choice of generator + $\sqrt{3} \,  \alpha_m \,  k_4$ on the Kerr string one gets a {non-extremal spinning magnetic one-brane.}
The five dimensional metric and gauge field take the form
\bea
ds^2 &=& \bar h \left[ - \xi^{-1} f (dt + \omega_{\phi} d \phi)^2 + ds^2_{base}\right] + \bar h ^{-2} \xi \left( dz + \hat A_{t} dt + \hat A_{\phi} d \phi \right)^2, \nonumber \\
A_{t} &=& \sqrt{3} \frac{c^2}{s^2} \frac{\beta_t}{\bar h},   \quad A_{z} = - \sqrt{3} \frac{c}{s} \frac{\beta_t}{\bar h}, \quad A_{\phi} = - 2 \sqrt{3} m c s x - \sqrt{3} (1-x^2) a \frac{ c }{ s^2} \frac{\beta_t}{\bar h} ~,
\label{magnetic}
\eea
where we have defined the following functions
\bea
\omega_{\phi} &=& \omega c^3, \qquad \beta_t = \frac{2 s^3  m\, a\, x}{\Sigma}, \\
h &=& c^2 - s^2 f, \qquad \xi = h^3 - \beta_t^2 f, \qquad g = 1 + \frac{1}{h^2 s^2 } \beta_t^2, \\
 \hat h &=& -s^2 f + c^2 g^{-1}, \qquad
\bar  h = \xi \hat h^{-1},  \\
\hat A_{t} &=& \frac{4  m^2 a^2  c^3 s^3 x^2}{\Sigma^2 \xi},  \\
\hat A_{\phi} &=&  \frac{a m s^3 (1-x^2)}{\xi \Sigma^2} \left[ 2 r \Sigma - 4 a^2 m x^2 +
 4 m s^2 (3 r^2 + 6 m r s^2 + 4 m^2 s^4) \right],
  \eea
and we have introduced
\be
s := \sinh \alpha_m, \qquad c := \cosh \alpha_m~.
\ee
The rest of the metric functions and the base metric are defined in equations (\ref{funcs}) and (\ref{base}) respectively.
The static limit of this solution is simply
\be
ds^2_{5} = H^{-1} \left( -\tilde f dt^2 + dz^2 \right) + H^2 \left( \frac{dr^2}{\tilde f} + r^2 d \Omega_2^2 \right),
\ee
with
\be
\tilde f = \left( 1 - \frac{2m}{r}\right), \qquad H = 1 + \frac{Q}{r}\: , \qquad Q = 2 m s^2 ,
\ee
and the gauge field is
\be
A_{\phi}= - 2 \sqrt{3} m \: s \, c \: x~.
\ee
From  the eleven dimensional point of view, we  recognize the static solution as representing a configuration of three, equally charged, intersecting M5 branes wrapped on a seven torus. The brane intersection is indicated in the following table
\begin{displaymath}
\begin{array}{c c c c c c c c c c c c}
 & t & z_1& z_2& z_3& z_4 & z_5 & z_6 & z & r & \theta & \phi \\
  M5 & \times & \times & \times & \times & \times & - & - & \times &  &  &  \\
  M5 & \times & \times & \times& - & - & \times & \times & \times &  &  &  \\
  M5 &  \times & - & - & \times & \times & \times & \times  & \times &  &  &
\end{array}
\end{displaymath}
The spinning solution (\ref{magnetic}) thus naturally represents the spinning intersection of this brane configuration with equal M5 charges. The spinning solution with three \emph{unequal} charges was recently obtained in \cite{Tanabe:2008vz} using boosts and string dualities. Upon setting the three charges equal, the magnetic solution of \cite{Tanabe:2008vz} reduces to (\ref{magnetic}) \footnote{The solution given in \cite{Tanabe:2008vz} had several typos that we have fixed.}.

We now provide an analysis of  physical properties of the boosted generalization of the spinning magnetic string (\ref{magnetic}).
We study the boosted string configuration because after bending it into a circle and for a specific value of the boost parameter (see equation (\ref{specific_mag}) below), it should give a doubly spinning
dipole black ring.
The boosted string configuration can be obtained by performing a boost (\ref{boost}) with the boost parameter $\sigma$ on (\ref{magnetic}). We take the new boosted $z$ coordinate to be along an $S^1$ with circumference $2 \pi R$, which allows us to write $z$ in terms of an angular coordinate $\psi$ defined in (\ref{angle}). Below we use $z$ and $\psi$ interchangeably.

It is easy to see that the magnetic solution (as well as its boosted sibling) also has a regular outer horizon
at $r= r_{+} :=  m + \sqrt{m^2 -a^2}$ of topology $R \times S^2$. In addition there is an inner horizon at $r= r_{-} := m - \sqrt{m^2 -a^2}$.
 The two horizons coincide when $a = m$, which defines the extremal limit.

The ADM stress tensor for the boosted solution is
\bea
T_{tt} &=& \frac{m}{2 } \left( 1+ \cosh^2 \sigma + 3 s^2   \right)~, \nn \\
T_{zz} &=& \frac{m}{2 } \left( \sinh^2 \sigma - 1 - 3 s^2   \right)~, \nn \\
T_{tz} &=& \frac{m}{2 }  \sinh \sigma \cosh \sigma~. \label{stress_mag}
\eea
The calculation of the mass, linear momentum, angular momentum, horizon area, and linear and angular velocities at the outer horizon is straightforward, though somewhat tedious, for this solution.  One finds
\bea
M &=&  2 \pi R \, \, T_{tt} = \pi m R \left( 1+ \cosh^2 \sigma + 3 s^2   \right)~, \label{masssmag}\\
P_{z} &=& 2 \pi R \, \, T_{tz} = \pi m R \sinh \sigma \cosh \sigma~, \\
J_{\phi} &=& 2 \pi R \: m a \: \left( c^3 \cosh \sigma + s^3 \sinh \sigma \right)~, \\
A_{\rom{H}} &=&  8 \pi^2 R \: \sqrt{\Xi}~, \label{aream} \\
v_{z} &=& \frac{a^2 (c^6 + s^6) \sinh \sigma \cosh \sigma  + c^3  s^3  (r_{+}^2 + 2 (a^2-2m^2 ) \cosh^2 \sigma) }{\Gamma }~,  \label{omegam}\\
\Omega_{\phi} &=& \frac{a \left( r_{-} c^3 \cosh \sigma  - r_{+} s^3 \sinh \sigma  \right)}{2 m \,  \Gamma}~, \label{omegaphi}
\eea
where we have defined two auxiliary functions $\Xi$ and $\Gamma$ to be
\bea
\Xi &=& (r_{+}^2 + a^2)^2 c^6 \cosh^2\sigma + (r_{-}^2 + a^2)^2 s^6 \sinh^2 \sigma -
  4 m^2 a^2 c^3 s^3  \sinh 2 \sigma~,  \\
\Gamma &=& a^2 c^6 \cosh^2 \sigma + a^2 s^6 \sinh^2 \sigma  + (a^2 - 2 m^2 ) c^3 s^3 \sinh 2 \sigma~.
\eea
Remarkably, for the \emph{un}-boosted solution (i.e., $\sigma = 0$ ) the $v_{z}$ linear velocity is non-zero
\be
v_{z} \big{|}_{\sigma = 0}
= - \frac{a^2}{r_+^2} \tanh^3\alpha_m,\label{omega0}
\ee
even though $P_{z}$ is zero! Since (\ref{omega0}) vanishes if either $a = 0$ or $\alpha_m = 0$, this is a combined effect of the internal rotation and the magnetic charge. Note that the orientation of the effect is dictated only by the sign of the magnetic charge and not by the sense of the internal rotation.  The effect can probably be attributed to the Chern-Simons coupling as it is similar to several peculiar frame dragging effects observed in five dimensional gravity
coupled to gauge fields with non-supersymmetric Chern-Simons couplings \cite{Kleihaus:2007kc}.

The temperature can be calculated from the surface gravity, and the result is
\be
T_{\rom{H}} = \frac{r_{+} - r_{-}}{4 \pi \sqrt{\Xi}}~.
\ee
As expected, $T_{\rom{H}} = 0$ for the extremal solution with $m = a$. The magnetic charge is
\be
Q_{\rom{M}} :=  \frac{1}{4 \pi}\int_{S^2} F =   -2\sqrt{3} \, m \, c \,  s~.
\ee

In addition, for studying thermodynamics of this solution, we define ADM tension ${\cal T}$ from the $T_{zz}$ component of the stress tensor,
\be
{\cal T} = - T_{zz} = \frac{m}{2} \left( 1 +  3 s^2   - \sinh^2 \sigma \right).
\label{tensionmag}
\ee

We next compute the chemical potential associated with the magnetic charge. Following \cite{Copsey:2005se} we work with regular gauge potentials in two patches: the north patch ($0 \leq \theta \leq \frac{\pi}{2}$ and $t$ constant) and the south patch  ($\frac{\pi}{2} \leq \theta \leq \pi$ and $t$ constant). We denote by $E$ the boundary between the two patches, that is, the surface of constant time $t$ and $\theta = \frac\pi 2$. Our gauge potentials satisfy the following boundary conditions
\begin{eqnarray}
A^{\rom{North}}_t &=& O(r^{-1}), \qquad A^{\rom{North}}_\phi = Q_{\rom{M}}  (x-1) + O(r^{-1}),\\
A^{\rom{South}}_t &=& O(r^{-1}), \qquad A^{\rom{South}}_\phi = Q_{\rom{M}}  (x+1) + O(r^{-1}).
\end{eqnarray}
With the choice of the constant $\Lambda$ such that
\begin{eqnarray}
\Lambda^{\rom{North}} &=& Q_{\rom{M}} \Omega_\phi ,\qquad \Lambda^{\rom{South}} = - Q_{\rom{M}}  \Omega_\phi~,
\end{eqnarray}
the quantity $A_\rho \xi^\rho + \Lambda$ is continuous across $E$. In reference \cite{Copsey:2005se} it was shown that the contributions to the first law coming from the surface terms on the surface $E$ is a term of the form  \be \Phi_{\rom{M}}\delta Q_{M}. \ee A general expression for the magnetic potential $\Phi_{\rom{M}}$ was also presented  in the Hamiltonian form. In the Lagrangian form it can be expressed as follows
\be
\Phi_{\rom{M}} := \frac{1}{8\pi } \int_{E} (d^{3}x)_{\mu\nu} \left(  2 \xi^\mu  F^{\alpha\nu} \partial_\alpha \phi  - \Omega_\phi F^{\mu \nu} \right)  + \frac{1}{8\sqrt{3}\pi}   \int_{E} F_{\alpha\beta} \partial_\gamma \phi (A_\rho \xi^\rho + \Lambda) dx^\alpha \wedge dx^\beta \wedge dx^\gamma \label{pot_magn}
\ee
where $(d^{3}x)_{\mu\nu} = \frac{1}{2\ 3!}\eps_{\mu\nu\alpha\beta\gamma} dx^\alpha \wedge dx^\beta \wedge dx^\gamma$. The first two contributions come from the Maxwell action and the last contribution comes from the Chern-Simons term.
For the boosted spinning magnetic string, only the Maxwell part of the magnetic potential is non-vanishing.
An explicit calculation gives the magnetic potential to be
\be
\Phi_{\rom{M}} =  -\frac{\sqrt{3} \pi m R }{2 \sqrt{m^2 - a^2}+2m \cosh 2\alpha_m } \left(1-a \, \Omega_\phi \cosh{\left(\alpha_m + \sigma\right)}\right) \sinh 2\alpha_m,
\label{phiM1}
\ee
which upon substituting the expression (\ref{omegaphi}) for $\Omega_{\phi}$ can be rewritten as
\be
\Phi_{\rom{M}} = - \frac{\sqrt{3} \, c \, s \, \pi \, R }{2} \: \frac{r_{+}^2 c^4 \cosh^2 \sigma + r_{-}^2 s^4 \sinh^2 \sigma - a^2 c s (c^2 + s^2) \sinh \sigma \cosh \sigma  }{ r_{+}^2 c^6 \cosh^2 \sigma + r_{-}^2 s^6 \sinh^2 \sigma - 2 a^2 c^3 s^3 \sinh \sigma \cosh \sigma }.
\label{phiM2}
\ee

It can be easily checked that the above expressions (\ref{masssmag})--(\ref{tensionmag}) and (\ref{phiM1}) reduce to the correct expressions for the boosted non-spinning magnetic one brane when $a =0$. {This boosted non-spinning one-brane describes the infinite radius limit of the singly spinning dipole ring of \cite{Emparan:2004wy}.}

A somewhat laborious calculation using these results then shows that the boosted black string satisfies the Smarr relation
\be
M = \frac{3}{2} \left( \frac{1}{4}T_{\rom{H}} A_{\rom{H}} + \Omega_\phi J_{\phi} + \Omega_{\psi} J_{\psi} \right)  + \frac{1}{2} {\cal T} (2 \pi R) + \frac{1}{2} \Phi_{\rom{M}} Q_{\rom{M}},
\ee
and the first law
\be
dM = \frac{1}{4} T_{\rom{H}} dA_{\rom{H}} + \Omega_{\psi}d J_{\psi} + \Omega_{\phi}d J_{\phi} +  \Phi_{\rom{M}} d Q_{\rom{M}} + 2 \pi  {\cal T} d R~,
\ee
where we have introduced the `angular' velocity and `angular' momentum \be\Omega_{\psi} = \frac{v_{z}}{R}, \qquad J_{\psi} = P_{z} R. \ee
For the pressureless solution ($T_{zz} =0$) the first law is exactly that of \cite{Copsey:2005se}. This hints to the possibility that the pressureless black string correctly describes the infinite radius limit of a five dimensional asymptotically flat black ring. When the magnetic charge and the internal spin are zero, the Smarr relation and the first law become exactly those of \cite{Kastor:2007wr}.

As also mentioned above, bending a boosted version of the magnetic string (\ref{magnetic}) into a circle should give a doubly spinning dipole black ring. An exact solution describing such a ring configuration is not known in the literature. It is likely that the exact  ring solution could be obtained by applying Yazadjiev solution generating technique \cite{Yazadjiev:2006hw, Yazadjiev:2006ew} to Pomeransky Sen'kov solution \cite{Pomeransky:2006bd}. However, to the best of our knowledge, such a construction has not yet been attempted. The Yazadjiev technique requires reducing the five dimensional theory to two dimensions; and therefore, in the case of five dimensional supergravity,  would require to work with the affine extension of the $G_2$ Lie algebra.

At any rate, motivated by considerations of \cite{Hovdebo:2006jy, Emparan:2007wm, Caldarelli:2008pz, Camps:2008hb, Emparan:2009cs}, here we simply study the boosted spinning magnetic string as a toy model for a thin doubly spinning dipole black ring.
We start by noting that the  stress tensor (\ref{stress_mag}) is identical to the one presented in \cite{Emparan:2004wy}, i.e., the internal spin does not enter in these components of the stress tensor.
Hence, the pressureless condition translates into the same specific value for the boost parameter as in \cite{Emparan:2004wy}
\be
\sinh^2 \sigma = 1 + 3 s^2. \label{specific_mag}
\ee
When $s\neq 0$ the boost is larger than in the neutral case. This observation is easily interpreted \cite{Emparan:2004wy} by noting that  sections of the ring at diametrically opposite ends, $\psi$ and $\psi + \pi$, have opposite orientations and therefore they attract each other via the 2-form field strength $F_{\mu \nu}$. As a result, a larger centrifugal repulsion is needed in order to achieve the mechanical equilibrium. It is interesting to contrast this situation with the electric solution described above -- for the electric solution a smaller centrifugal repulsion is needed, as in that case the opposite ends on the ring repel each other. Substituting (\ref{specific_mag}) in equations (\ref{masssmag})--(\ref{tensionmag}) and (\ref{phiM1}), one can extract certain important information about the balanced doubly spinning dipole black ring.

\subsection{$k_5$: NUT charge}
Acting with the choice of generator $ + \frac{1}{2} \, \alpha \, k_5$ on the Kerr string one gets the Kerr-Taub-NUT string.
Upon performing the following coordinate and parameter changes
$ r \rightarrow m + r - m \cos \alpha, \: M = m \cos \alpha, \: N = m \sin \alpha, $
the final metric is in the standard Boyer-Lindquist coordinates.

\subsection{$k_6$: KK monopole}
Applying   $+ \frac{1}{2} \, \alpha \, k_6$ on the Kerr string one gets the spinning thermally excited KK-monopole solution of \cite{Rasheed:1995zv, Larsen:1999pp}.

\section{Conclusions}
\label{disc}
In this paper we have explored the $G_2$ solution generating technique for five dimensional minimal supergravity. This technique requires reducing the theory on two commuting Killing directions.
Upon dimensional reduction, the bosonic equations of motion reduce to those of three dimensional gravity coupled to a non-linear sigma model.  When the reduction
is performed over two spacelike Killing directions one obtains the $G_2/SO(4)$ coset model. On the other hand, when the reduction is performed over one
timelike and one spacelike Killing direction one obtains the $G_2/(SL(2,\RR) \times SL(2,\RR))$ coset model. In section \ref{dimred} we reviewed this dimensional
reduction, treating both cases simultaneously.

The $G_2$ symmetries of the coset model can be used to construct new solutions by applying group transformations on seed solutions.
In section \ref{cartannilpotent} we considered the action of the Cartan and $N_+$ subgroups of $G_2$
on a general seed solution and showed that they act as scaling
and gauge transformations respectively. Most interesting is the non-linear action of
the subgroup $SL(2,\RR) \times SL(2,\RR)$. Its action on a general seed solution is very complicated and not illuminating, so we considered its action only on  a particular
seed solution, namely, on the five dimensional
Kerr string.
We expect that the six generators of $SL(2,\RR) \times SL(2,\RR)$ charge a general seed solution with Kaluza-Klein asymptotics in essentially the same way as they charge the Kerr string.
Their action on the Kerr string is given in Table \ref{actiononKerr}. {The action of the appropriate pseudo-compact group, in particular, how it rotates charges of BPS black holes, for a closely related $N=2$ $d=4$ supergravity theory is currently under investigation \cite{Axel}.}
\begin{table}[!h]
\begin{center}
\begin{tabular}{|c|c|}
\hline
$k_1$ & boost \\ \hline
$k_2$ & electric charge \\ \hline
$k_3$ & generates nothing \\ \hline
 $k_4$ & magnetic charge \\ \hline
 $k_5$ & NUT charge\\ \hline
 $k_6$ & KK monopole \\
\hline
\end{tabular}
\end{center}
\caption{Action of the $SL(2,\RR) \times SL(2,\RR)$ subgroup of $G_2$ on the five dimensional Kerr string.}
\label{actiononKerr}
\end{table}

We expect that in any coset model  the decomposition of symmetry generators into Cartan, nilpotent and (pseudo-)compact generators
plays a similar role as it does in the $G_2$ coset model.
More precisely, we predict
 that in any coset model obtained via dimensional reduction, e.g., the $E(8)/K(E(8))$ model of eleven dimensional supergravity, all generators belonging to the Cartan and nilpotent subgroups act as scaling and gauge transformations, while the compact or pseudo-compact subgroup act as charging transformations.

Using the $SL(2,\RR) \times SL(2,\RR)$ action we obtained a spinning electric and a spinning magnetic black string. These solutions were also recently obtained in string theory in reference \cite{Tanabe:2008vz}. Our method is more efficient compared to \cite{Tanabe:2008vz} in obtaining these solutions in minimal supergravity.
  We  analyzed physical properties of these black strings and studied their thermodynamics. We also explored their relation to black rings.

Thanks to the efficiency of the $G_2$ method, the black string describing the infinite radius limit
of the conjectured most general black ring of five dimensional minimal supergravity can also be constructed.
Here we give a brief outline of how this construction proceeds.
Let us start by recalling that such a black string (without imposing the pressureless condition) must have five independent parameters: mass, internal spin, boost, smeared electric charge, and magnetic one-brane charge.
At first sight, one might hope that by the successive action of $k_2$ and $k_4$ on the Kerr string one would generate a four parameter dyonic black string. However, in doing so one also generates a Lorentzian NUT charge and hence Dirac-Misner strings singularities\footnote{In fact, viewing the $k_2$ and $k_4$ actions as efficient ways of doing sequences of
boosts and string dualities, such Dirac-Misner strings are expected to arise \cite{Elvang:2003mj, Elvang:2004xi}.}.
To get around this difficulty, one might start
with a seed solution that already has a NUT parameter, viz., the Kerr-Taub-NUT string, with the hope
that by tuning the extra initial parameter one will be able to cancel
the final NUT charge. It turns out that even doing this is not sufficient! One can indeed
cancel the NUT charge, but, in the process
of successively applying $k_2$ and $k_4$ on the Kerr Taub-NUT string one
also generates a KK
monopole charge. To cancel the KK monopole  charge one needs to  apply another $G_2$ transformation, $k_6$, that
adds another parameter. Appropriately tuning this and the initial NUT parameter one gets a smooth
dyonic string. Finally, applying a boost, one generates the requisite five parameter black string. The solution and its physical properties will be presented in a separate publication \cite{monster}.

A natural continuation of this work is to extend our analysis to black rings and to solutions with more general horizon topologies. A step in this direction was taken in \cite{Bouchareb:2007ax} where the authors considered the action of $k_2$ on the Pomeransky Sen'kov solution \cite{Pomeransky:2006bd}.  A four parameter electrically charged solution was constructed, but it suffers from Dirac-Misner string singularities. It was argued in \cite{Bouchareb:2007ax} that, in principle, using $G_2$ dualities it should be possible to generate a six parameter unbalanced black ring, which should lead to a four parameter non-singular black ring \cite{Bouchareb2}.

An even farther reaching generalization consists in reducing the $G_2$ sigma model on one more Killing direction. This would allow one to act with the affine extension of $G_2$, $G_2^+$, on solutions of five dimensional minimal supergravity. This line of investigation has been extensively explored in recent years for the gravitational sub-sector of this theory, i.e., for five dimensional vacuum gravity. It has led to a great recent progress in our understanding of stationary black holes with two rotational Killing vectors (see \cite{Emparan:2008eg} for a review). The main reason for this success is that after reduction to two dimensions, vacuum five dimensional gravity is completely integrable. As a result, powerful solution generating techniques are available. It is expected that the full minimal supergravity reduced on three commuting Killing vectors also leads to a completely integrable sigma model.
If this is the case, then acting with $G_2^+$ might lead  a way to a complete classification of stationary axisymmetric solutions of this theory.

\subsection*{Acknowledgements}
We thank Marc Henneaux for suggesting this project and for comments on a draft. We thank Riccardo Argurio, Glenn Barnich, Thibault Damour, St\'ephane Detournay, Pau Figueras, Daniel Persson, and Philippe Spindel for discussions.  We are especially grateful to Roberto Emparan and Axel Kleinschmidt for their constant interest in our project and for feedbacks and advice on several of the issues addressed in this paper. We acknowledge Keith Copsey and Gary Horowitz for great discussions about the magnetic potential. This work is supported in part by the US National Science Foundation under Grant No.~PHY05-55669, and by funds from the University of California, the Universit\'e Libre de Bruxelles, and  the International Solvay Institutes.  SdB is grateful to the European Commission for its financial support though the grant PIOF-GA-2008-220338 based at the Universit\'e Libre de Bruxelles, Belgium. EJ is a FRS-FNRS bursar. EJ and AV were supported by IISN - Belgium convention
4.4505.86 and by the Belgian Federal Science Policy Office through the Interuniversity Attraction Pole P5/27.

\appendix
\section{Explicit representation of $\mf{g}_{2(2)}$ \label{expl_rep}}
{\color{black}In this appendix we give the representation of $\mf{g}_{2(2)}$ that we use:
\beq \small{
h_1 = \left(
\begin{array}{lllllll}
 \frac{1}{\sqrt{3}} & 0 & 0 & 0 & 0 & 0 & 0 \\
 0 & -\frac{1}{\sqrt{3}} & 0 & 0 & 0 & 0 & 0 \\
 0 & 0 & \frac{2}{\sqrt{3}} & 0 & 0 & 0 & 0 \\
 0 & 0 & 0 & 0 & 0 & 0 & 0 \\
 0 & 0 & 0 & 0 & -\frac{2}{\sqrt{3}} & 0 & 0 \\
 0 & 0 & 0 & 0 & 0 & \frac{1}{\sqrt{3}} & 0 \\
 0 & 0 & 0 & 0 & 0 & 0 & -\frac{1}{\sqrt{3}}
\end{array}
\right)
 \, , \hspace{1cm}
h_2 = \left(
\begin{array}{lllllll}
 1 & 0 & 0 & 0 & 0 & 0 & 0 \\
 0 & 1 & 0 & 0 & 0 & 0 & 0 \\
 0 & 0 & 0 & 0 & 0 & 0 & 0 \\
 0 & 0 & 0 & 0 & 0 & 0 & 0 \\
 0 & 0 & 0 & 0 & 0 & 0 & 0 \\
 0 & 0 & 0 & 0 & 0 & -1 & 0 \\
 0 & 0 & 0 & 0 & 0 & 0 & -1
\end{array}
\right) \, , } \no \eeq
\beq
\small{
e_1 = \left(
\begin{array}{lllllll}
 0 & 0 & 0 & 0 & 0 & 0 & 0 \\
 0 & 0 & 1 & 0 & 0 & 0 & 0 \\
 0 & 0 & 0 & 0 & 0 & 0 & 0 \\
 0 & 0 & 0 & 0 & 0 & 0 & 0 \\
 0 & 0 & 0 & 0 & 0 & 1 & 0 \\
 0 & 0 & 0 & 0 & 0 & 0 & 0 \\
 0 & 0 & 0 & 0 & 0 & 0 & 0
\end{array}
\right) \, ,\hspace{1cm}
e_2 = \left(
\begin{array}{lllllll}
 0 & \frac{1}{\sqrt{3}} & 0 & 0 & 0 & 0 & 0 \\
 0 & 0 & 0 & 0 & 0 & 0 & 0 \\
 0 & 0 & 0 & \frac{2}{\sqrt{3}} & 0 & 0 & 0 \\
 0 & 0 & 0 & 0 & \frac{2}{\sqrt{3}} & 0 & 0 \\
 0 & 0 & 0 & 0 & 0 & 0 & 0 \\
 0 & 0 & 0 & 0 & 0 & 0 & \frac{1}{\sqrt{3}} \\
 0 & 0 & 0 & 0 & 0 & 0 & 0
\end{array}
\right) \, .} \no \eeq
\beq
\small{
f_1 = \left(
\begin{array}{lllllll}
 0 & 0 & 0 & 0 & 0 & 0 & 0 \\
 0 & 0 & 0 & 0 & 0 & 0 & 0 \\
 0 & 1 & 0 & 0 & 0 & 0 & 0 \\
 0 & 0 & 0 & 0 & 0 & 0 & 0 \\
 0 & 0 & 0 & 0 & 0 & 0 & 0 \\
 0 & 0 & 0 & 0 & 1 & 0 & 0 \\
 0 & 0 & 0 & 0 & 0 & 0 & 0
\end{array}
\right)\, ,\hspace{1cm}
f_2 = \left(
\begin{array}{lllllll}
 0 & 0 & 0 & 0 & 0 & 0 & 0 \\
 \frac{1}{\sqrt{3}} & 0 & 0 & 0 & 0 & 0 & 0 \\
 0 & 0 & 0 & 0 & 0 & 0 & 0 \\
 0 & 0 & \frac{1}{\sqrt{3}} & 0 & 0 & 0 & 0 \\
 0 & 0 & 0 & \frac{1}{\sqrt{3}} & 0 & 0 & 0 \\
 0 & 0 & 0 & 0 & 0 & 0 & 0 \\
 0 & 0 & 0 & 0 & 0 & \frac{1}{\sqrt{3}} & 0
\end{array}
\right)\, .} \no \eeq
The representation of the other generators follows from the commutation relations (\ref{genepos}) and the definitions (\ref{newbasis}).}

\section{Killing symmetries}
\label{Killsym}

In this appendix we show that using finite dimensional hidden symmetries one cannot generate a solution that has less number of commuting Killing symmetries than that of a seed solution. Suppose that the seed solution admits $N \geq 2$ commuting Killing symmetries: $\chi_a$, $a=1,\dots, N$. Therefore, $$\cL_{\chi_a}g_{\mu\nu}= 0 \quad \mbox{and} \quad \cL_{\chi_a}A_{\mu}= d\epsilon_a$$  for some gauge  parameters $\epsilon_a$. When we dimensionally reduce along the orbits of two Killing vectors, the three dimensional metric and the scalar fields are all invariant under the action of $\chi_a$'s. Indeed, one can choose locally $N$ coordinates on the spacetime to be $\hat\chi_a$
such that $\chi_a = \frac{\partial}{\partial \hat \chi_a}$.  The fields $g_{\mu\nu}$ and $A_\mu$ then do not depend on $\hat\chi_a$ and from the definitions \eqref{metric5}-\eqref{pot5}-\eqref{dualfields} we immediately see that all the three dimensional fields also do not depend on $\hat\chi_a$'s.

The new scalar fields obtained from an action of $G_2$ are then also invariant under $\chi_a$'s, because they are constructed from certain (non-linear) combinations of the original fields. The dual gauge fields obtained after integrations, in \eqref{dualfields} or \eqref{matrixN}, are also invariant under $\chi_a$'s possibly up to gauge transformations. Indeed, we have $\cL_{\chi_a}(\star \cM^{-1} d\cM) =\cL_{\chi_a} (d \mathcal N)  = 0$ by hypothesis. Therefore, $d( \cL_{\chi_a} \mathcal N )= 0$ and so $\cL_{\chi_a} \mathcal N = d\epsilon$. The result then follows from the definition of the dual gauge fields.

As a duality transformation admits an inverse, we have shown that solutions related by dualities always have the same amount of commuting Killing symmetries. This precludes the use of the $G_2$, or for that matter any finite dimensional hidden symmetries, to generate solutions with less number of commuting Killing symmetries than the ones that are already known.

\section{Asymptotic analysis}
\label{asym}

It is important to note that in general the asymptotic structure of the seed spacetime is not preserved by transformations generated by coset symmetries.  This appendix is devoted to analyze the generators of $G_2$ that preserve asymptotic flatness and Kaluza-Klein asymptotics. See also \cite{Clement:2008qx, Bouchareb2}.

\subsection{Asymptotic flatness}
\label{symmatrixM}
We follow Giusto and Saxena \cite{Giusto:2007fx} to find generators that preserve five dimensional asymptotic flatness by  focusing on the asymptotic limit of the \emph{symmetric form}\footnote{Reference \cite{Giusto:2007fx} uses a construction based on the matrix $\tilde{\cM}$ defined as $\tilde{\cM}: =  \cV^T \eta \cV $
where $\eta $ is a matrix invariant under $SL(2,\RR)\times SL(2,\RR)$. The coset Lagrangian in terms of the matrix $\tilde{\cM}$ is again given by the expression (\ref{cosetLag}), that is,
$ \cL_{coset} = {1 \over 8}\mathrm{Tr}(d\tilde \cM^{-1}d\tilde
\cM) $. The matrix $\tilde \cM$ transforms as $\tilde \cM
\rightarrow g^T \tilde \cM g$ when $\cV \rightarrow k \cV g$.
Moreover, the matrix $\eta$ allows to pass from the transposition to
the generalized transposition via $g^t = \eta g^\sharp \eta^{-1}$.
The matrix $\eta$ in our representation is given by $ \eta  = c $ diag$(-1,1,-1,2,-4,4,-4)$, where $c$ is a constant that we choose to be $c=-1/2$. The relationship between our matrix
$\cM$ and $\tilde \cM$ is $\cM = \eta^{-1} \tilde \cM$. Therefore, when $\cV \rightarrow k \cV g$ and $\cM \rightarrow g^\sharp
\cM g $,  $\tilde \cM \rightarrow g^T \tilde \cM g$.}
 $\tilde \cM$ of the matrix $\cM$.
For Minkowski spacetime we denote the  matrix $\tilde \cM$ by $\tilde \cM_{\RR^{1,4}}$.  To find the generators preserving asymptotic flatness we use the simple criteria of \cite{Giusto:2007fx} that asymptotically the matrix $\tilde \cM$ for a general spacetime should approach the asymptotic limit $\tilde \cM_{\RR^{1,4} }^\infty$ of flat space matrix $\tilde \cM_{\RR^{1,4}}$. This criteria is not sufficient to discard certain pathological spacetimes that approach flat space at the boundary only locally.

In order to get the  matrix $\tilde \cM_{\RR^{1,4}}$, one should, as in the pure gravitational case \cite{Giusto:2007fx}, compactify Minkowski space on two Killing vectors. Following \cite{Giusto:2007fx} we do this in the spherical polar coordinates
\be
ds^2 = -dt^2 + dr^2 + r^2 \left( d \theta^2 + \sin^2 \theta d\phi^2 + \cos^2 \theta d \psi^2 \right)
\ee
 with the choice
\beq
z_4 &=& \ell (\phi-\psi)  \quad \mbox{or}  \quad z_4 = \ell \left(\phi + \psi \right) \\
z_5 &=&t~.
\eeq
With these choices $\tilde \cM_{\RR^{1,4}}$ and its asymptotic form $\tilde \cM_{\RR^{1,4} }^\infty$
read
\beq \small{ \tilde{\cM}_{\RR^{1,4}}Ê= \left( \begin{array}{ccccccc}
{ 2 \ell^2 \over r^2} &  0 & 0 & 0 & 0& -1&0 \\
 0& -{ 2 \ell^2 \over r^2} & 0 & 0 & 0 & 0 & 1 \\
  0 & 0 & 1/2 & 0 & 0 & 0 & 0 \\
 0 &0 &0 &-1 &0 &0 & 0 \\
0& 0& 0& 0& 2& 0& 0 \\
-1& 0& 0& 0& 0& 0& 0\\
0& 1& 0& 0& 0& 0& 0\\
\end{array}
\right) \hspace{1cm}Ê\tilde \cM_{\RR^{1,4} }^\inftyÊ= \left( \begin{array}{ccccccc}
0 &  0 & 0 & 0 & 0& -1&0 \\
 0& 0& 0 & 0 & 0 & 0 & 1 \\
  0 & 0 & 1/2 & 0 & 0 & 0 & 0 \\
 0 &0 &0 &-1 &0 &0 & 0 \\
0& 0& 0& 0& 2& 0& 0 \\
-1& 0& 0& 0& 0& 0& 0\\
0& 1& 0& 0& 0& 0& 0\\
\end{array}
\right). }\no
\eeq
The isotropy group of $\tilde \cM_{\RR^{1,4} }^\infty$, i.e., $\{ g \in GL(7,\RR) \arrowvert g^T \tilde \cM_{\RR^{1,4} }^\infty  g = \tilde \cM_{\RR^{1,4} }^\infty \}$, is not really transparent in this basis.
After a suitable basis change $P$, the matrix $\tilde \cM_{\RR^{1,4} }^\inftyÊ$ reads $\tilde \cM_{\RR^{1,4} }^{\infty \, \prime}=P^T \, \tilde \cM_{\RR^{1,4} }^\inftyÊ\, P$ = diag$(-1,1,1,-1,1,1,-1)$. From this form it immediately follows that the isotropy group of  $\tilde \cM_{\RR^{1,4} }^\infty$ is contained in $SO(3,4)$. The generators of $G_{2}$ that belongs to the isotropy group are the following,
\beq
h_2\, , \, \, e_3 + e_4\, , \, \, e_1 + e_5\, , \, \, k_2 \, , \, \, 2 e_3 - k_3 + k_4\, , \, \,   -2 e_1 + k_1 - k_5 \, .
\eeq
The three gravitational generators $h_2$, $e_1+e_5$ and $-2e_1 + k_1 -k_5$ correctly reproduce the commutation relations of the expected  $SO(2,1)$ found for vacuum five dimensional gravity \cite{Giusto:2007fx}.  In appendix \ref{giusto} we show how the results of \cite{Giusto:2007fx} are embedded in our formalism.

\subsection{Black string asymptotics}
 One can perform a similar analysis for the Kaluza-Klein asymptotics  $\RR^{3,1}\times S^1$. It goes along the same lines as for the asymptotically flat case. We consider a reduction on time $z_5=t$ and the `string direction' $z_4=z$. The main result is that the entire pseudo-compact group
$\tilde K \subset G_2$ preserves the Kaluza-Klein asymptotics  $\RR^{3,1}\times S^1$. It is interesting to contrast this with the asymptotically flat case where only three combinations out of the six generators of the pseudo-compact algebra preserve asymptotic flatness. The action of $G_2$ on black strings is therefore richer than on five dimensional black holes.

\section{$SL(3,\RR) \subset G_{2(2)}$: the gravitational sector}
\label{giusto}

In this appendix, we compare the dimensional reduction ansatz of \cite{Giusto:2007fx} with ours. This allows us to embed the results of \cite{Giusto:2007fx} in minimal supergravity.

When the five dimensional gauge field is set to zero, minimal supergravity reduces to vacuum gravity. This tells us that the hidden symmetry  $SL(3,\RR)$ of vacuum five dimensional gravity is part of $ G_{2(2)}$.  Indeed, $\cH_{grav}\simeq \mf{sl}(3,\mathbb R)$ is a subalgebra of $\mf{g}_{2}$. It is generated by the elements
$(h_1,h_2,e_1,e_5,e_6, f_1,f_5,f_6)$. The simple roots of $\cH_{grav}$ are in one-to-one correspondence with the scalars
obtained from the reduction \eqref{eq:redLagr} of the gravitational sector of the supergravity Lagrangian.

In this appendix we would like to understand how the $3\times 3$ Maison matrix \cite{Maison:1979kx}  $\chi$, used in \cite{Giusto:2007fx}, is embedded in our $7 \times 7$  matrix $\tilde \cM$, defined in appendix \ref{symmatrixM}. See also \cite{Bouchareb:2007ax,Clement:2008qx}.
To this end let us start  by comparing our dimensional reduction ansatz (\ref{metric5}) with the one used in \cite{Giusto:2007fx}
\begin{eqnarray}
ds^2_{5 \, ours} &=& e^{\frac{1}{\sqrt 3}\phi_1+\phi_2}ds^2_3 +\eps_2 e^{\frac{1}{\sqrt 3}\phi_1-\phi_2}(dz_4 + \cA^2_{(1)})^2\nonumber\\
&&+\eps_1 e^{-\frac{2}{\sqrt 3}\phi_1}(dz_5+\cA^2_{(0)1}dz_4+\cA^1_{(1)})\, , \\
ds^2_{5 \, GS} &=& \lambda_{ab} (d\xi^a + \o^a{}_i dx^i) ( d\xi^b+ \o^b{}_j dx^j) + { 1\over \tau } ds_3^2 \, .
\label{metricGS}
\end{eqnarray}
With the choice $\xi^1 = z_5, \, \xi^2 = z_4$, a comparison of $44, 45$ and $55$ components of the two ansatzes give the $2 \times 2$ matrix $\lambda_{ab}$ used in (\ref{metricGS}) in terms of our fields:
\beq
\lambda  = \left(
\begin{array}{ll}
\epsilon_1 e^{-{2\over \sqrt{3}} \phi_1} & \epsilon_1 e^{-{2\over \sqrt{3}} \phi_1}  \cA^2_{(0) 1} \\
\epsilon_1 e^{-{2\over \sqrt{3}} \phi_1}  \cA^2_{(0) 1}   &
 \epsilon_2 e^{{1\over \sqrt{3}} \phi_1- \phi_2} +  \epsilon_1 e^{-{2\over \sqrt{3}} \phi_1}  \cA^2_{(0) 1}  \cA^2_{(0) 1}  \, , \\
\end{array}
\right) \, ,
\eeq
and thus $\tau := \mbox{det} \lambda =  e^{-{1\over \sqrt{3}} \phi_1- \phi_2}$. Comparing  the $ij$ components
\be
\lambda_{ab} \o^a{}_i \o^b{}_j = \epsilon_2 e^{{1\over \sqrt{3}} \phi_1- \phi_2}(\cA^2_{(1)})^2_{ij} + \epsilon_1 e^{-{2 \over \sqrt{3}}\phi_1 } (\cA^1_{(1)})^2_{ij} \, ,
\ee
we get the one--forms $\omega^{b}$'s in terms our fields
\be
\o^1 =\cA^1_{(1)} - \chi_1 \cA^2_{(1)} \, ,  \qquad \o^2 = \cA^2_{(1)}~.
\ee
Finally, to compare the scalars $V_a$'s used in \cite{Giusto:2007fx} with our axions, we must dualize the one--forms $\omega^{b}$'s  via the relation
\beq
dV_a = - \tau \,  \lambda_{ab}  \, \star_3 d \o^b \, . \label{dualgs}
\eeq
The easiest way to do this is to write $d \o^b$ in terms of axions $\chi_5 $, $\chi_6$ and the dilatons. To achieve this, we use the truncation of equation (\ref{dualfields}) obtained by
setting the three dimensional fields $\chi_{2}, \chi_{3}$ and $A_{(1)}$  to zero. The fields $\chi_{2}, \chi_{3}$ and $A_{(1)}$ correspond to the five dimensional gauge field $A^{5}_{(1)}$ which we need to set to zero in order to get to vacuum five dimensional gravity.
Equation (\ref{dualfields}) then reduces to
\begin{eqnarray}
\eps_1 e^{-\vec\alpha_5 \cdot \vec\phi} \star  \cF_{(2)}^1& \equiv &G_{(1)5} = d\chi_5~, \\
\eps_ 2 e^{-\vec\alpha_6 \cdot \vec\phi}\star  \cF_{(2)}^2  &\equiv &G_{(1)6} = d\chi_6- \chi_1 d\chi_5~,
\end{eqnarray}
{\color{black}where
$ \cF_{(2)}^1= d\cA^1_{(1)}+ \cA^2_{(1)} \wedge d \chi_1$ and $\cF_{(2)}^2=d\cA^2_{(1)} $.} Using these relations $d \o^b$'s are readily expressed in terms of $\chi_5 $, $\chi_6$ and the dilatons. The result is
{\color{black}
\begin{eqnarray}
\star \, d \o^2 &=& \star \cF^2_{(2)} =  \eps_ 2 e^{\vec\alpha_6 \cdot \vec\phi}(d\chi_6- \chi_1 d\chi_5 ) \, , \\
\star \, d \o^1 &=& \star \cF^1_{(2)} - \chi_1 \star \cF^2_{(2)} =  \eps_1 e^{\vec\alpha_5 \cdot \vec\phi}d\chi_5 - \eps_ 2 e^{\vec\alpha_6 \cdot \vec\phi}(d\chi_6- \chi_1 d\chi_5 ) \, . \end{eqnarray}}
From equation (\ref{dualgs}) it now follows that  \be V_1 = - \chi_5~, \qquad \: V_2 = - \chi_6~.\ee

Using a different matrix representation\footnote{The representation is given by the matrix $Z$ given on page 1656 of  reference \cite{Gunaydin:1973rs}. {\color{black}We denote the representation matrices of \cite{Gunaydin:1973rs} by bars over the generators to distinguish them from our representation. }} for $G_{2(2)}$ than ours and the following coset representative,
\beq
{\color{black}e^{ \frac{1}{4} \bar h_1 \left(\sqrt{3} \phi _1-\phi _2\right)+\frac{1}{4} \bar h_2 \left(\phi _1+\sqrt{3} \phi _2\right)}
e^{-\bar e_5  \chi _6} e^{- \bar f_1 \chi _1} e^{- \bar e_6\chi _5}} \eeq
we obtain
\beq
\tilde \cM  = \left(
\begin{array}{ccc}
\chi^{-1} & 0 & 0 \\
0  & \chi & 0 \\
0 & 0 & 1
\end{array}
\right) \, ,
\eeq
where
\beq \nonumber
\chi = \left(
\begin{array}{ccc}
 -e^{-\frac{2 \phi _1}{\sqrt{3}}} \left(1+e^{\sqrt{3} \phi _1+\phi _2} \chi _5{}^2\right) & -e^{-\frac{2 \phi _1}{\sqrt{3}}} \left(\chi _1+e^{\sqrt{3} \phi _1+\phi _2} \chi _5 \chi _6\right) & -e^{\frac{\phi _1}{\sqrt{3}}+\phi _2} \chi _5 \\
 -e^{-\frac{2 \phi _1}{\sqrt{3}}} \left(\chi _1+e^{\sqrt{3} \phi _1+\phi _2} \chi _5 \chi _6\right) & e^{-\frac{2 \phi _1}{\sqrt{3}}-\phi _2} \left(-e^{\phi _2} \chi _1{}^2+e^{\sqrt{3} \phi _1} \left(1-e^{2 \phi _2} \chi _6{}^2\right)\right) & -e^{\frac{\phi _1}{\sqrt{3}}+\phi _2} \chi _6 \\
 -e^{\frac{\phi _1}{\sqrt{3}}+\phi _2} \chi _5 & -e^{\frac{\phi _1}{\sqrt{3}}+\phi _2} \chi _6 & -e^{\frac{\phi _1}{\sqrt{3}}+\phi _2}
\end{array}
\right).
\eeq
This $\chi$ is identical to the one used in \cite{Giusto:2007fx}.

We are now in position to  compare the action of the generators $N_\alpha$, $N_\beta$, $N_\gamma$, $M_\alpha$, $M_\beta$, $M_\gamma$, $D$ defined respectively in (2.36)-(2.37) and (2.24) in \cite{Giusto:2007fx} with our generators. All we have to do is to compute the action of these generators on the fields $\chi_1$, $\chi_5$, $\chi_6$ and $\phi_1$, $\phi_2$ using the dictionary for the fields given above and find the analogous transformations in our notation. The comparison of generators is given in Table~\ref{compaSax}. We have checked that the action of $M_\alpha$ on the five dimensional Schwarzschild black hole with the choice $\xi^1 = t$, $\xi^2 = \ell \left(\psi+\phi\right)$ generates the doubly spinning Myers-Perry black hole with equal rotation parameters in the two rotation planes as in \cite{Giusto:2007fx}.
\begin{table}[!hdt]
\begin{tabular}{|c|cc|}
\hline
\text{Generators of \cite{Giusto:2007fx}} & \text{Our generator} & \text{Interpretation of \cite{Giusto:2007fx} on asymptotics} \\
\hline
$D$ & $e^{-\frac \pi 4 k_6 + \frac \pi 2 e_6}$ & Change asymptotic behavior of $\tilde{\cal M}$ from $\mathbb R^{4,1}$  to $\mathbb R^{3,1} \times S^1$ \\
$N_\alpha$ & $e^{\alpha k_1}$ & Preserve $\mathbb R^{3,1} \times S^1$ asymptotic \\
$N_\beta$ & $e^{\frac \pi 4 k_6 - \frac \pi 2 e_6} e^{\beta h_2}e^{-\frac \pi 4 k_6 + \frac \pi 2 e_6}$ & Preserve $\mathbb R^{3,1} \times S^1$ asymptotic \\
$N_\gamma$ & $e^{\gamma k_5}$ & Preserve $\mathbb R^{3,1} \times S^1$ asymptotic \\
$M_\alpha =D^T N_\alpha D$ & $e^{-\frac \pi 4 k_6 +\frac \pi 2 e_6} e^{\alpha k_1} e^{\frac \pi 4 k_6 - \frac \pi 2 e_6}$ & Generate rotation while preserving $\mathbb R^{4,1}$ up to a large diffeo \\
$M_\beta = D^T N_\beta D$ & $e^{\beta h_2}$ & Large diffeomorphism in $\mathbb R^{4,1}$ \\
$M_\gamma = D^T N_\gamma D$ & $e^{-\frac \pi 4 k_6 + \frac \pi 2 e_6} e^{\gamma k_5} e^{\frac \pi 4 k_6 - \frac \pi 2 e_6}$ & Generate rotation while preserving $\mathbb R^{4,1}$ up to  a large diffeo \\
\hline
\end{tabular}\caption{Summary of the interpretation of the action of the generators of \cite{Giusto:2007fx}. The composition of two transformations is in reverse order as compared to \cite{Giusto:2007fx} since in our conventions the group $G_2$ acts as \eqref{transfM}. }\label{compaSax}
\end{table}

\end{document}